\documentclass[preprint,12pt,authoryear]{elsarticle}
\usepackage{float}
\usepackage{amssymb}
\usepackage{textgreek}
\usepackage{upgreek}
\usepackage{url}
\usepackage[utf8x]{inputenc}
\usepackage[T1]{fontenc}
\usepackage{amsmath}
\usepackage{amsfonts}
\usepackage{graphicx}
\graphicspath{{images/}}
\usepackage{fancyhdr} 
\usepackage{vmargin}
\usepackage{verbatim}
\usepackage{hyperref}
\usepackage{multirow}
\usepackage{appendix}
\usepackage{placeins}
\usepackage{array}
\usepackage{xcolor}
\definecolor{green}{RGB}{0,180,0}

\usepackage{todonotes}
\reversemarginpar

\begin{document}

\begin{frontmatter}

\title{Experimental and Monte Carlo Simulation Studies to Characterize a Low-Pressure Gas-Based Positive-Ion Detector with a Nanometer-Equivalent Sensitive Volume\tnoteref{t1}}
\tnotetext[t1]{Prepared for submission to RADMEAS}

\author[label1,label2,label3]{Victor Merza\corref{cor1}}
\ead{victor.merza@tecnico.ulisboa.pt}
\cortext[cor1]{Corresponding author}
            
\author[label4,label5]{Aleksandr Bancer}

\author[label3]{Vladimir Bashkirov}
    
\author[label1,label2]{Ana Belchior}

\author[label6]{Beata Brzozowska}

\author[label1,label2]{João F. Canhoto}

\author[label7,label8]{Piotr Gasik}

\author[label5]{Jaroslaw Grzyb}

\author[label1,label2]{Khaled Katmeh}

\author[label4,label5]{Marcin Pietrzak}

\author[label4]{Antoni Ruciński}

\author[label3]{Reinhard Schulte}

\affiliation[label1]{organization={Centro de Ciências e Tecnologias Nucleares, Instituto Superior Técnico, Universidade de Lisboa},
            addressline={Estrada Nacional 10 (km 139,7)}, 
            city={Bobadela LRS},
            postcode={2695-066}, 
            country={Portugal}}
\affiliation[label2]{organization={Departamento de Física, Instituto Superior Técnico, Universidade de Lisboa},
            addressline={Av. Rovisco Pais 1}, 
            city={Lisboa},
            postcode={1049-001},
            country={Portugal}}
\affiliation[label3]{organization={Department of Basic Science, Division of Biomedical Engineering Sciences, Loma Linda University},
            addressline={201 Mortensen Hall, 11085 Campus St}, 
            city={Loma Linda},
            postcode={92350}, 
            state={California},
            country={United States of America}}
\affiliation[label4]{organization={Cyclotron Centre Bronowice, Institute of Nuclear Physics Polish Academy of Sciences},
            addressline={Radzikowskiego 152}, 
            city={Kraków},
            postcode={31-342}, 
            country={Poland}}
\affiliation[label5]{organization={Radiological Metrology and Biomedical Physics Division, National Centre for Nuclear Research},
            addressline={ul. Andrzeja Sołtana 7}, 
            city={Otwock Świerk},
            postcode={05-400},
            country={Poland}}
\affiliation[label6]{organization={Biomedical Physics Division, Institute of Experimental Physics, Faculty of Physics, University of Warsaw},
            addressline={5 Pasteur Street}, 
            city={Warsaw},
            postcode={02-093}, 
            country={Poland}}
\affiliation[label7]{organization={GSI Helmholtzzentrum f\"ur Schwerionenforschung GmbH (GSI)},
            addressline={Planckstr. 1}, 
            city={Darmstadt},
            postcode={64291}, 
            country={Germany}}
\affiliation[label8]{organization={Physikalisches Institut, Ruprechts-Karls-Universität Heidelberg},
            addressline={Im Neuenheimer Feld 226}, 
            city={Heidelberg},
            postcode={69120},
            country={Germany}}

\begin{abstract}
Compact low-pressure gas-based positive-ion detectors offer a promising approach to measuring particle track structure at the nanometer scale.
These detectors use electric fields to collect and multiply positive ions produced by the incident radiation in mm-sized holes in dielectric materials, achieving a nm-equivalent spatial resolution of the localization of ionization events when scaled to liquid water at unit density.

In this work, a detector prototype featuring a 1.5~mm-diameter hole in a dielectric plate was experimentally characterized.
Ion collection and multiplication were achieved by applying a negative high potential to the glass cathode 0.5~mm below the hole, assisted by a low drift field above the plate.
A grounded readout electrode with a 0.8~mm hole covers the hole to prevent interactions of collected ions with the hole walls.
The experimental setup was further modeled with Geant4-DNA and Garfield++ for deeper insight.
By combining the measurements with the Monte Carlo simulations, a per-ion signal formation probability in the low single-digit percentage range was determined.
Further, the signal formation mechanism, commonly attributed to ion-impact ionization within the hole, was critically examined.
Ion-induced secondary electron emission from surfaces was considered a possible contributor to, or the primary mechanism of, signal formation that has not been addressed in detail in previous studies of this detector concept.
Despite the extensive measurements and simulations, the available evidence did not reveal which physical processes governed signal formation or whether either mechanism dominates.
We discuss key limitations in the current understanding of this detector class and propose future targeted measurements and simulations to clarify the signal formation mechanism and guide the development of compact positive-ion detectors and nanodosimeters.
\end{abstract}

\begin{keyword}
Nanodosimetry \sep particle track structure \sep GEM \sep THGEM \sep Geant4 \sep Geant4-DNA \sep Garfield++
\end{keyword}

\end{frontmatter}

\section{Introduction}
Nanodosimetry aims to predict DNA damage frequency and complexity based on the track structure of ionizing particles, which shows the nanometric spatial distribution of excitations and ionizations in tissue \citep{rucinski,conte2023,Faddegon}.
Ionizations are considered the dominant contributors to radiobiological effectiveness, either by generating diffusing radicals that induce indirect DNA damage or by directly ionizing the DNA molecule.
In particular, when more than four ionizations occur in a clustered pattern within a DNA molecule or within a few nanometers of it, the probability of inducing irreparable or lethal lesions increases significantly \citep{rucinski}.
The ionization cluster size, which is the number of ionizations produced by a primary particle and its secondaries in the considered nanometric volume, e.g., a DNA segment covering about 10-20 base pairs, is associated with the DNA lesion cluster size.
When adjacent DNA lesions induced by an ionizing particle occur within a separation of no more than 10 base pairs, they are defined as a lesion cluster \citep{rucinski}.
Nanodosimetric quantities specifying the spatial distribution of ionization events are measurable.
Thus, by specifying the nanometric pattern of ionization clustering, nanodosimetric quantities can be used as an alternative descriptor of radiation quality that can be validated experimentally \citep{mietelska2024,conte2017}.

\begin{figure}[thbp]
    \centering
    \includegraphics[width=0.7\linewidth]{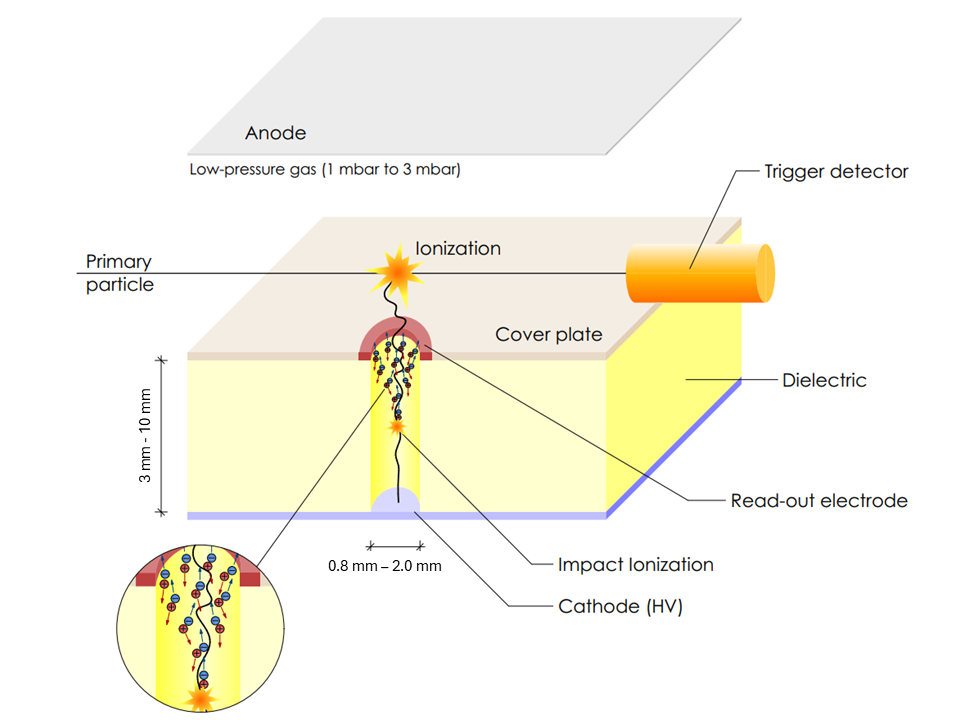}
    \caption{This schematic illustrates the hypothesized ion‑impact ionization (III) mechanism of the signal formation. Alternative mechanisms such as ion-induced secondary electron emission (IISEE) from the cathode may also contribute or even dominate signal formation.}
    \label{fig_working_principle}
\end{figure}

Nowadays, since no technology allowing measurements of the track structure in tissue or tissue-equivalent materials at the nanometer level has been developed so far, the measurement of nanodosimetric quantities is only feasible in low-pressure gases, and thus in nanometer-equivalent sensitive volumes (SV).
There are only a few nanodosimeters developed in the 1990s and early 2000s that are still operational.
These include the ion counter detector \citep{garty2002,garty2002b,bashkirov2006,bashkirov20091,hilgers2015,hilgers2017,hilgers2019,hilgers2022}, the LNL Startrack Counter \citep{denardo2002,denardo2002b,denardo2007,conte2010,conte2011,conte2012,Grosswendt2014,conte2014,selva2020,conte2023}, and the Jet Counter \citep{pszona2000,bantsar,pietrzak2018,bancer2020}.
Most recent research aimed to design compact detectors with nanometer-equivalent resolution \citep{bashkirov20092,casiraghi2014,casiraghi2015,vasi2016,FIRE,kempf2025,kempf20252,merza2026}, combining the working principles of thick gas electron multipliers (THGEM), and resistive plate chambers (RPC) operated at low pressure and in reverse polarity.
These compact detectors collect the positive ions produced by incident radiation from the SV and guide them into holes with diameters ranging from 0.8~mm to 2~mm in dielectric plates with thicknesses of 3~mm to 10~mm.
In the hole, the collected ions generate a self-amplified signal (Fig.~\ref{fig_working_principle}).
Ion collection is achieved through a combination of a low drift field provided by an anode above the dielectric plate and a high-voltage cathode, which is attached to the bottom side of the plate and provides a strong electric field inside the hole.
Due to the low diffusion of the ions, a nanometer-equivalent spatial resolution of the starting position of the ions and thus the accurate 3D-localization of the ionizations can be achieved.

Although the efficiency of these detectors is not yet sufficient for practical applications, they are of great interest due to their potential for miniaturization.
Advanced miniaturized nanodosimeters could provide a tool in medical physics or radiation protection for the validation of nanodosimetry-based treatment plans or the characterization of radiation quality.

The working principle proposed for previously developed detectors, including the TIDe \citep{bashkirov20092}, FIRE \citep{FIRE}, FIRE-V2 \citep{kempf20252}, and the detector presented in \citep{vasi2016}, assumes that the signal is induced by ion-impact ionization (III) of the gas molecules inside the holes by the accelerated positive ions \citep{bashkirov20092}.
However, experimental evidence demonstrating this mechanism as the dominant cause of signal formation is lacking.
Alternatively, ion-induced secondary electron emission (IISEE) from the cathode or the hole surfaces may contribute to, or even dominate, signal formation.
Distinguishing between these mechanisms is essential because they imply different approaches to optimizing the detector performance.

The primary goal of this study was to characterize the operation of a compact low-pressure gas-based detector prototype through experimental testing and Monte Carlo simulations.
By comparing the measured detector response with simulation results, we examined the experimental observations in the context of III and IISEE as potential signal formation mechanisms.
We also identified limitations in the modelling of the detector response and outlined further experimental and computational investigations needed to distinguish between these mechanisms.

\section{Materials and Methods}

\subsection{Experimental Methods}\label{sec_expmeth}

\subsubsection{Detector Setup}
\begin{figure}[thbp]
    \centering
    \includegraphics[width=\linewidth]{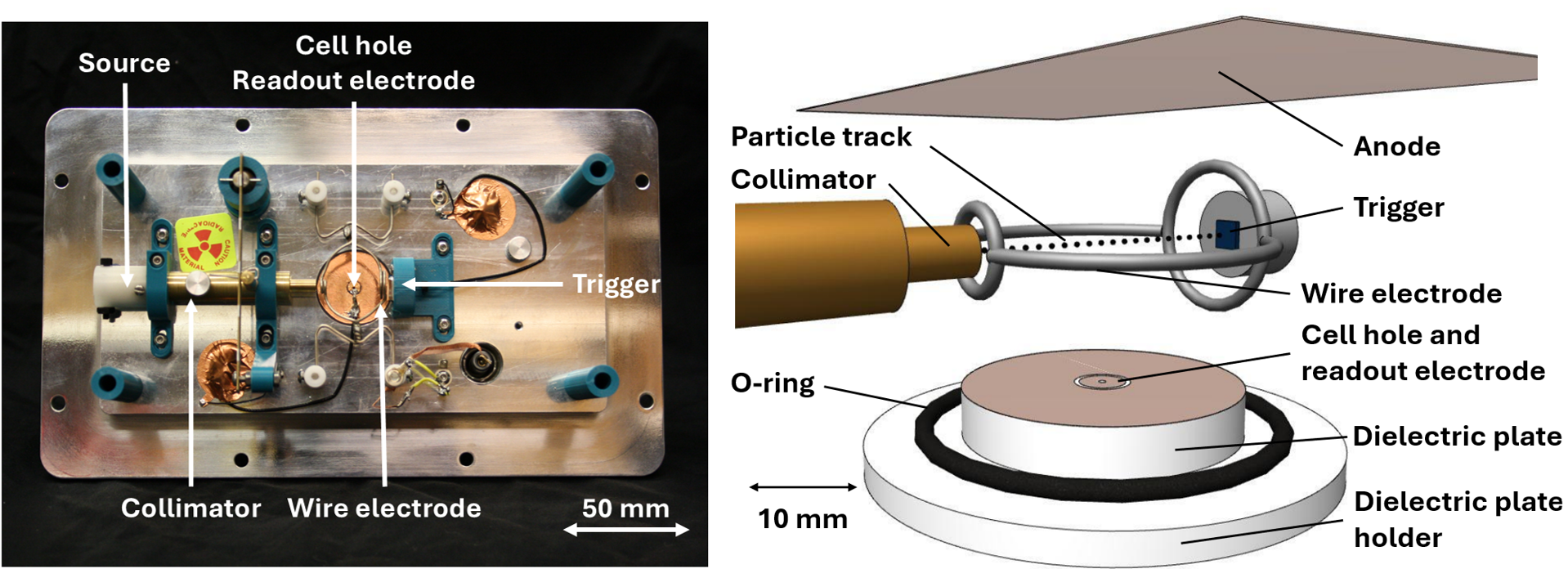}
    \caption{Left: Top view of the detector interior attached to the lid of the low-pressure chamber. Right: Enlarged schematic view of the detector setup.}
    \label{fig_detector}
\end{figure}

The detector prototype used in the measurements is shown in Fig.~\ref{fig_detector}, on the left.
The detector comprises a low-pressure chamber with a dielectric plate mounted in an opening in the chamber lid.
The plate is attached to a holder that fits into the opening, where an O-ring between the holder and the chamber lid provides the vacuum seal (Fig.~\ref{fig_detector}, right).
This design was inspired by the detector assembly presented in \citep{FIRE,kempf20252}.
A hole, where the detector signal is produced, is mechanically drilled into this plate.
The design of this detector aims to investigate the signal formation mechanism rather than to measure the ionization cluster size distribution (ICSD), which is the frequency distribution of the ionization cluster size produced in the SV.

An Am-241 source is positioned on the lid within the low-pressure gas volume, together with a trigger detector that registers the primary alpha particles.
Also, electrodes providing the drift field for the positive ions produced in the drift region are mounted to the lid.

The 3~mm thick dielectric plate made of acrylic (polymethyl methacrylate, PMMA) features a single 1.5~mm diameter hole.
On the top surface of the plate, which faces the low-pressure gas drift region traversed by the primary particles, a readout electrode made of copper foil is attached using non-conductive adhesive.
The readout electrode, which features an outer diameter of 5~mm and a central circular opening with a diameter of 0.8~mm, is positioned concentrically with the hole.
A key difference from other similar prototypes is that the readout electrode has a smaller diameter than the dielectric, which prevents collected ions from hitting the hole walls.
The remaining copper-foil-covered surface of the dielectric plate is not in contact with the readout electrode.
Both structures are held at the same bias of 0~V.
A rectangular copper anode plate with dimensions of 80~mm~$\times$~40~mm, positioned 20~mm above the readout electrode, provides the drift field.
At a distance of 10~mm above the dielectric plate, halfway from the anode, a silver-cladded copper wire ring (30~mm in diameter) is suspended from the anode holders (white PTFE spacers that hold and fix the wire electrode, visible in Fig.~\ref{fig_detector}, left) to ensure a uniform electric field distribution.
A 1~mm collimated alpha-particle beam (see below for more details) traverses the chamber at 10~mm above the dielectric plate.
To allow unobstructed passage of the beam through the silver-clad copper wire electrode, which is positioned at the same height, the wire forms a small loop around the beam path on either side (Fig.~\ref{fig_detector}, on the right).
The central beam axis of the primary particle beam traversing the volume between the anode and the readout electrode passes directly above the center of the dielectric plate hole.

\begin{figure}[thbp]
    \centering
    \includegraphics[width=0.7\linewidth]{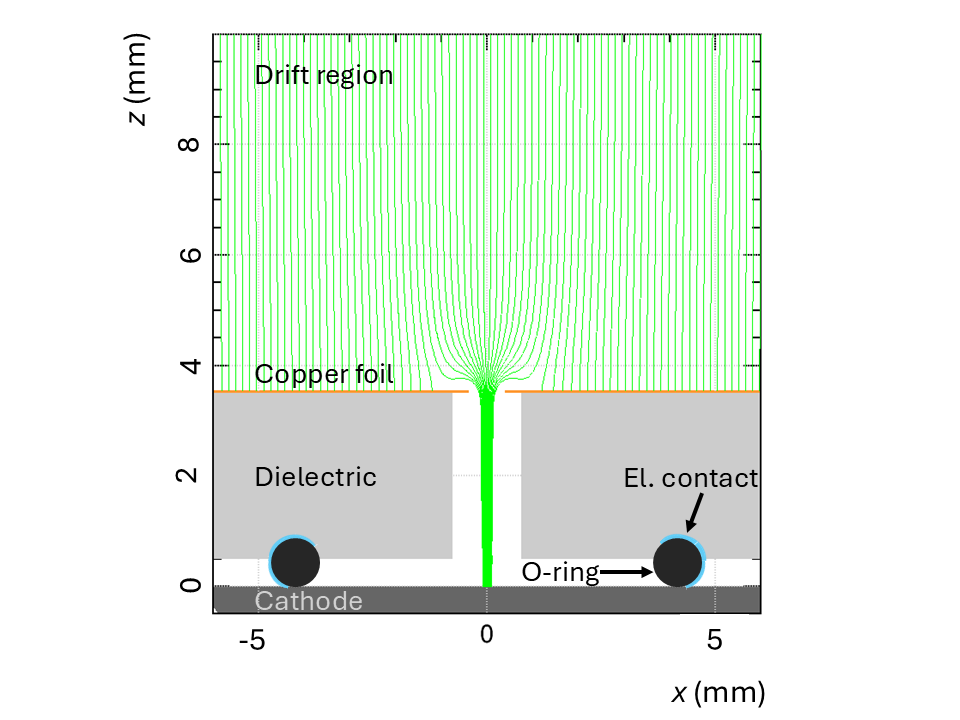}
    \caption{Schematic of the detector in the $x$–$z$ plane, with electric field lines calculated from equidistant starting points along the central beam axis at $z=13.5$ mm. The cathode and anode potentials are $-800\,\rm V$ and $30\,\rm V$, respectively. The anode is located at $z=23.5$ mm (not shown).}
    \label{fig_configuration}
\end{figure}

A 0.7~mm thick low-resistive glass cathode measuring 12~mm~$\times$~12~mm is attached to the bottom side of the dielectric plate.
The bulk resistivity of the glass is of the order of $10^{10}~\Omega$cm \citep{wang2019,wang2010}.
An O-ring with an outer diameter of 9.6~mm and an inner diameter of 6~mm is seated on the air side of the dielectric plate, with the cathode placed on top, providing sufficient sealing of the assembly.
Although there is a groove to embed this O-ring in the dielectric, a small gap of approximately 0.5~mm arises between the cathode and the plate (Fig. \ref{fig_configuration}).
To ensure direct electrical contact between the cathode and the dielectric plate, two contacts made of conductive ink are applied on opposite sides of the air side of the O-ring.
The high voltage was directly connected to the air-facing surface of the cathode, which was covered with an aluminum foil.

The Am-241 source (Eckert \& Ziegler, catalog number AM1A2100U) emitting alpha particles with a mean energy of 4.6~MeV is attached to a brass collimator with a length of 80~mm at a distance of 95~mm from the hole axis.
The beam-defining aperture of the collimator has a diameter of 1~mm.

A PIN photodiode with a square sensitive area of 3.2~mm~$\times$~3.2~mm, positioned 15~mm from the hole axis, registers primary particles at a rate of about 2.2~Hz and is used as the trigger detector.

The data acquisition system includes a digital Picoscope 3206D oscilloscope (Pico Technology Ltd., UK), picking up the signals, which are analyzed retrospectively.
It is required to attenuate the detector signal produced in the hole to a level suitable for an ORTEC 9301 preamplifier (ORTEC, Oak Ridge, Tennessee, USA) by terminating its input line with an additional 50~$\Omega$ to ground.
Further, a 6~dB attenuator is used right before the preamplifier's input.
The trigger detector signals are amplified via a CAEN 1425 preamplifier (C.A.E.N. S.p.A., Viareggio, Italy). 
Both preamplifiers are connected to the oscilloscope input channels via 50~$\Omega$ terminators.

The gas system, providing a continuous gas flow in the low-pressure chamber of the detector, consists of a 248A flow control valve (MKS Inc., Andover, MA, USA), an MKS Type 250 pressure/flow controller, an MKS Type 626 Baratron manometer, and a manual precision valve to adjust the outflow to the vacuum pump.

The measurements were performed in propane gas at room temperature, $(24\pm1)^\circ$C, at pressures of 1~mbar and 2~mbar.
The cathode potential ranged from $-600$~V to $-1000$~V, and was incrementally increased in 50~V steps.
The anode potential was fixed at 30~V, and the silver-clad copper wire ring between the anode and the dielectric plate was at 15~V.
The voltages were applied using an ORTEC 710 power supply.

At the propane gas pressures used in this work, the corresponding $pd$ (gas pressure $p$ times gap length $d$) values are approximately $0.35\,\mathrm{Pa\,m}$ at 1~mbar and $0.70\,\mathrm{Pa\,m}$ at 2~mbar.
Accordingly, the detector operates below the Paschen minimum at 1~mbar and near the minimum at 2~mbar, with the minimum for propane reported at $\approx 0.67\ \mathrm{Pa\,m}$ \citep{heylen1975}.

\subsubsection{Characterization of Detector Performance} \label{sec_efficiency}
The performance of the detector was characterized by the signal yield $\eta$, which is the ratio of the number of triggers with a subsequent detector signal in timely coincidence to the total number of triggers.
Only one or no detector signal was recorded within the time window of $t=200$~\textmu s after each trigger.
The time distribution of the delays between the maxima of the trigger signals and the following detector signals within $t$, labeled time-to-signal, was recorded and fitted with a Gaussian function.
It was established, as a convention, that a detector signal occurred in timely coincidence with a previous trigger signal if it was recorded within $\Delta t_\mathrm{arr}$.
The time window $\Delta t_\mathrm{arr}$ was defined as $\pm3~\sigma$ around the maximum of the Gaussian fit.

The number $N$ of registered triggers, the number $N_\mathrm{coinc}$ of detector signals that occurred within $\Delta t_\mathrm{arr}$, and the number $N_\mathrm{no~coinc}$ outside of $\Delta t_\mathrm{arr}$, but within $t$, was determined.
To ensure conservatism in the calculation of the signal yield, $N_\mathrm{no~coinc}$ was excluded from the calculation of the signal yield.
The signal yield $\eta$ was then calculated as follows:
\begin{equation} \label{equ_efficiency}
\eta = \frac{N_\mathrm{coinc}}{N}
\end{equation}

\subsection{Detector Simulations}
The computational modeling of the conducted experiments included:
\begin{itemize}
    \item Radiation transport simulations were performed using the condensed-history Monte Carlo code Geant4 \citep{geant4_1,geant4_2,geant4_3}, version 11.0.3, to simulate the transport and energy deposition of the incident radiation in the detector geometry. Geant4-DNA \citep{G4DNA1,G4DNA2,G4DNA3,G4DNA4,G4DNA5}, an extension of Geant4 including low-energy interaction processes such as ionizations relevant for nanodosimetric quantities, was activated exclusively in the low-pressure gas volume to simulate the particle track structure and the associated ionization patterns. This included the SV, located within the gas volume, where the track structure was of particular interest. Since the Geant4-DNA physics models used in this study do not apply to the detector's solid materials, the remaining detector volumes were not simulated with Geant4-DNA. The resulting ionization positions in the gas volume were used as initial positions for the ion drift trajectory simulations described below.
    \item Electrostatic field simulations using the finite-element-method software Elmer \citep{elmerfem}, version 9.0.
    \item Ion drift trajectory simulations using the detector simulation toolkit Garfield++ \citep{garfieldpp}.
\end{itemize}
The computational methods for the detector simulations are only briefly described here as they have been described and discussed in more detail in our previous publication \citep{merza2025}.

\subsubsection{Track Structure Simulation}\label{methods_track_structure}
\begin{figure}[thbp]
    \centering
    \includegraphics[width=0.7\linewidth]{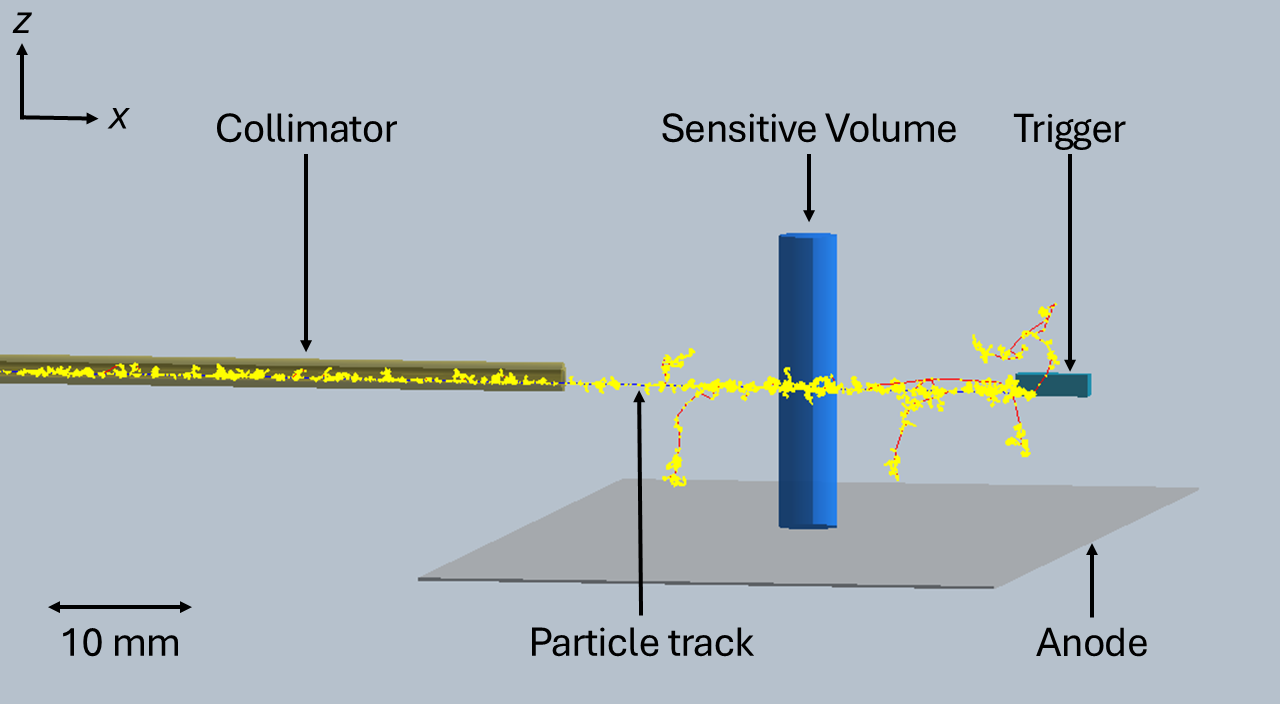}
    \caption{Simplified Geant4 model of the experimental setup. All components visible in the figure are located within the low-pressure gas volume. The collimator is represented only by the central gas-filled hole and a thin surrounding brass layer. The walls of the low-pressure chamber are not shown for better visibility.}
    \label{fig_G4_model}
\end{figure}

The simplified Geant4 model included the walls of the low-pressure chamber made of aluminum, an infinitely thin disk-shaped monoenergetic 4.6 MeV alpha source attached to the collimator, the trigger detector, and the copper anode (Fig.~\ref{fig_G4_model}).
Interior detector parts that were assumed not to influence the radiation transport, e.g., holders or electrodes outside of the beam, were not included in this model.

Using the \texttt{G4EMLivermorePhysics} list, the radiation transport was simulated in all volumes except for the low-pressure gas.
Geant4-DNA option 4 was chosen for the track structure simulation and the determination of the ICSD and the mean cluster size ($M_1$) in the SV (see Sec. \ref{sec_ion_drift}).
The positions of the ionizations in the drift region between the anode and the dielectric plate were recorded.
Since there were no available cross-section data of alpha particles in propane available in Geant4-DNA, a density scaling procedure \citep{grosswendt1,grosswendt2} was applied to substitute the propane gas with liquid water, taking into account the different mean free path lengths with respect to ionization for these two materials:
\begin{equation}\label{equ_scaling}
(D\rho)_{\mathrm{water}}=(D\rho)_{\mathrm{propane}} \cdot \frac{(\lambda\rho)_\mathrm{water}^{\mathrm{ion}}(Q)}{(\lambda\rho)_\mathrm{propane}^{\mathrm{ion}}(Q)}
\end{equation}
$(D\rho)_{\mathrm{\mathrm{propane}}}$ is the mass per area of propane at the corresponding pressure applied in the experiment, and $Q$ is the radiation quality.
The ratio of the mean free path lengths of water and propane with respect to ionization, $(\lambda\rho)_\mathrm{water}^{\mathrm{ion}}/(\lambda\rho)_\mathrm{propane}^{\mathrm{ion}} \approx~1.45$ \citep{hilgers2022,Bug2013}, is nearly energy-independent for alpha particles at the energies applied in this work.

\subsubsection{Electric Field Simulation}
For the calculation of the electrostatic fields with Elmer, a simplified geometrical model of the active detector part consisting of the cathode, dielectric plate, readout electrode, drift region, and anode was modeled.
The influence of the collimator, the detector walls, and the trigger detector on the electric field was not taken into account.
The relative permittivity of the acrylic was set to 2.7.

\subsubsection{Ion Drift Simulation and SV Definition}\label{sec_ion_drift}
Garfield++ simulations of the ion drift trajectories of single-ionized propane molecules (C$_3$H$_8^+$) in the electrostatic fields were performed using ion mobility values published in \citep{monte_carlo_model}.
For each setup, i.e., the chosen combination of gas pressure and cathode potential, the ion drift trajectories were calculated from the points of ionization simulated with Geant4-DNA (see above).
Ion starting positions were restricted to a cylindrical region, the sensor region, with a diameter of 10~mm and a height of 20~mm, extending from the top surface of the dielectric plate to the anode.
About $10^6$ starting ions were simulated, both with and without ion diffusion activated in Garfield++.
Diffusion was modeled using the macroscopic approach implemented in the \texttt{AvalancheMC} class, which statistically accounts for the cumulative effect of microscopic ion–neutral collisions.
The ion trajectory simulation was terminated when the ion reached any solid part of the detector, such as the readout electrode, the surrounding copper foil, or the cathode, and the coordinates of its endpoint were recorded.
An ion was considered collected if it passed through the opening of the hole.
The frequency distribution and the mean value $N_\mathrm{coll}$ of the number of collected ions per primary particle were recorded.
The duration for an ion to drift from its point of origin to the cathode is referred to as the ion arrival time.

To define the SV of the detector, which is required for the calculation of an ICSD using Geant4-DNA, the radial ion collection efficiency was determined.
The sensor region was divided into 300 concentric radial bins of width 0.05~mm.
For each bin, the ion collection efficiency was calculated as the ratio of ions collected to ions initially placed in that radial bin.
Consistent with our previous work \citep{merza2025}, which showed weak dependence of the ion collection efficiency on the vertical start position of the ion above the dielectric plate, efficiencies were averaged over the vertical coordinate within the sensor region to yield a radial profile.
This profile was interpolated using a cubic spline.
The SV was then defined as a cylinder whose radius corresponds to the radial distance at which the interpolated collection efficiency falls to 1/$e$.

\subsection{Estimation of the Signal Formation Probability}
The signal formation probability $p_\mathrm{s}$ for each collected ion was estimated from the measured signal yield $\eta$, i.e., the fraction of $N$ triggers that created a signal (Sec.~\ref{sec_efficiency}), and the number $n_i$ of collected ions for the $i^{\rm th}$ trigger ($i=1,...,N$), simulated with Garfield++ (Sec.~\ref{sec_ion_drift}).
In a first approximation, $p_\mathrm{s}$ can be inferred from $1-\eta$, i.e., the fraction of triggers that did not produce a signal, as
\begin{equation}\label{eq_ps}
1-\eta = \frac{1}{N} \sum_{i=1}^{N} (1-p_\mathrm{s})^{n_i}
\end{equation}
Eq.~\ref{eq_ps} implicitly shows the relationship between the signal formation probability $p_\mathrm{s}$ and signal yield $\eta$, which was solved numerically using the Python function \texttt{fsolve} to determine the value of $p_\mathrm{s}$.

\section{Results} \label{results}
\subsection{Track Structure and Ion Collection from the SV}\label{sec_res_track}
\begin{figure}[thbp]
    \centering
    \includegraphics[width=0.7\linewidth]{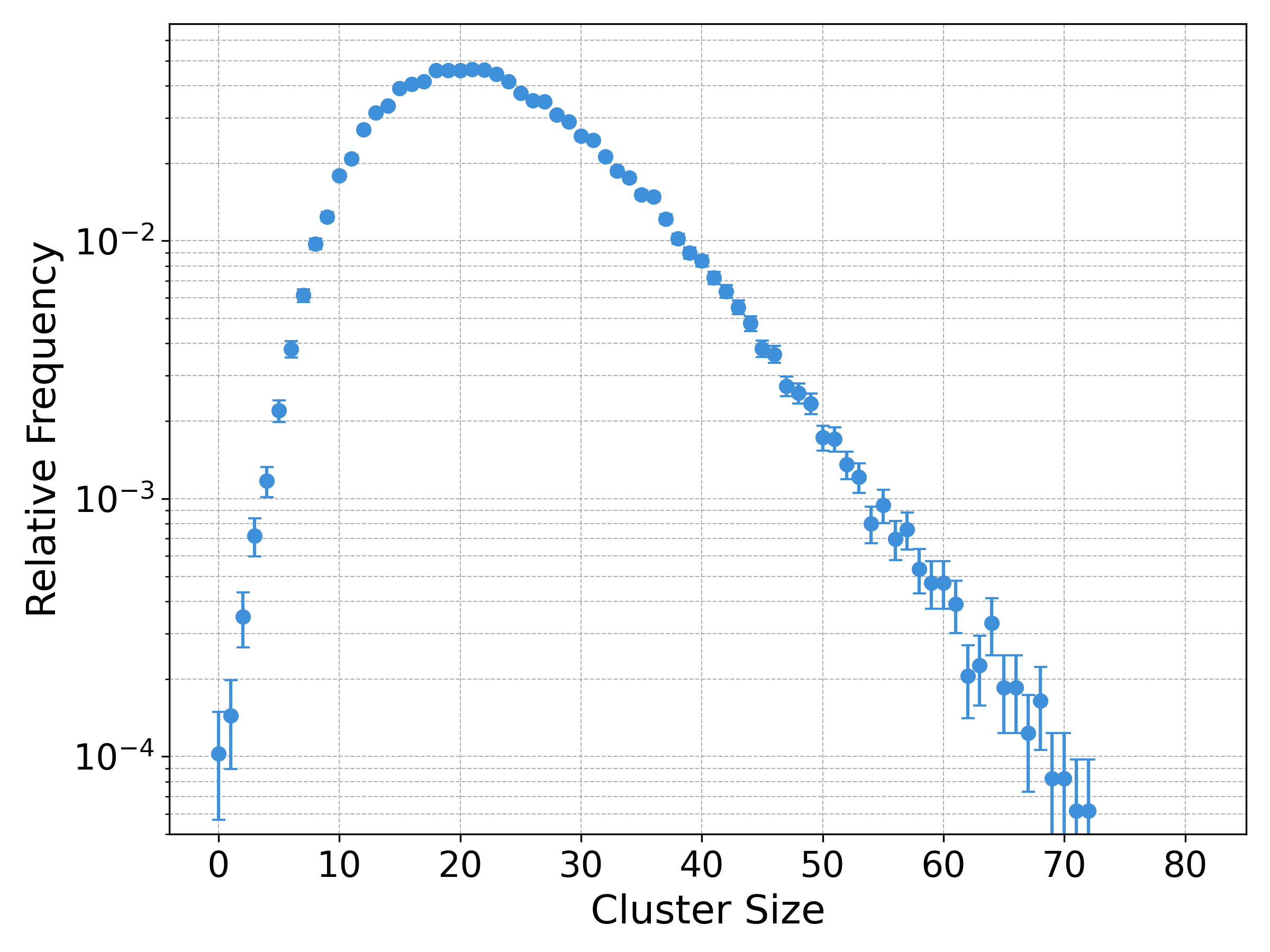}
    \caption{Geant4-DNA simulated ICSD produced in the SV with a diameter of 3.09~mm at a propane gas pressure of 1~mbar. For this plot, $2\cdot 10^5$ primary particles were simulated.}
    \label{fig_ICSD}
\end{figure}

Fig.~\ref{fig_ICSD} shows an example for the Geant4-DNA simulated ICSD produced by the 4.6~MeV alpha particles in the SV at 1~mbar propane gas pressure for $-800$~V cathode potential with an SV diameter of 3.09 mm.
This ICSD has a maximum at a cluster size of around 20 and rapidly falls off for higher and smaller cluster sizes, respectively.
The relative probability for zero ionizations produced in the SV is lower than $5\cdot10^{-4}$ for all configurations.

In Appendix A, Fig.~\ref{fig_collection_efficiency}, (a) and (b), the simulated ion collection efficiency is shown as a function of the radial distance from the hole center, averaged over the height of the drift region of 20~mm, for 1~mbar and 2~mbar propane gas pressure, respectively.
The collection efficiency approximately resembles a sigmoidal decrease with the radial distance from the center of the hole.
The influence of the electric field generated by the cathode, which penetrates through the hole in the readout electrode, results in broader radial collection efficiencies and thus in larger SV sizes for higher cathode potentials.
Tab.~\ref{tab_SV_sizes} lists the SV sizes, $M_1$-values derived from the ICSDs, and the Garfield++ simulated mean number $N_\mathrm{coll}$ of ions collected in the hole per incident primary alpha particle for each detector configuration.
Equivalent SV sizes in nm correspond to the size of the SV when scaled to liquid water with unit density according to Eq.~\ref{equ_scaling}.
$M_1$ and $N_\mathrm{coll}$ gradually increase as the cathode potential is varied from $-600$~V to $-1000$~V, with slightly more efficient ion collection at higher cathode potentials.
At 2~mbar, while the SV diameters $d_\mathrm{SV}$ in the propane gas are approximately the same as for 1~mbar, both $M_1$ and $N_\mathrm{coll}$ are slightly more than double the value at 1~mbar.
Note that $N_\mathrm{coll}$ represents the mean simulated number of collected ions in the static electric field.

\begin{table}
    \centering
\begin{tabular}{l|l|l|l|l|l}
    $p$ (mbar) & $U_\mathrm{Cath}$ (V) &  $d_\mathrm{SV}$ (mm) & $d_\mathrm{SV}^\mathrm{equ}$ (nm) & $M_1$ & $N_\mathrm{coll}$\\ \hline
    \multirow{9}{*}{1}
        & $-600$ & 2.81 & 6.90 & 20.87(8) & 16.82(7) \\
        & $-650$ & 2.93 & 7.20 & 22.00(9) & 17.51(7) \\
        & $-700$ & 2.94 & 7.23 & 22.13(9) & 18.08(7) \\
        & $-750$ & 3.00 & 7.37 & 22.68(9) & 18.70(7) \\
        & $-800$ & 3.09 & 7.58 & 23.46(9) & 19.29(7) \\
        & $-850$ & 3.11 & 7.63 & 23.65(9) & 19.83(8) \\
        & $-900$ & 3.17 & 7.78 & 24.21(9) & 20.37(8) \\
        & $-950$ & 3.23 & 7.93 & 24.78(9) & 20.94(8) \\
        & $-1000$ & 3.31 & 8.13 & 25.55(9) & 21.43(8) \\ \hline
    \multirow{9}{*}{2}
        & $-600$ & 2.82 & 13.88 & 48.66(24) & 38.94(18) \\
        & $-650$ & 2.91 & 14.29 & 50.46(24) & 40.50(19) \\
        & $-700$ & 2.95 & 14.48 & 50.10(25) & 41.79(20) \\
        & $-750$ & 3.03 & 14.91 & 53.09(25) & 43.36(20) \\
        & $-800$ & 3.08 & 15.16 & 54.14(25) & 44.74(20) \\
        & $-850$ & 3.12 & 15.34 & 54.96(26) & 46.01(21) \\
        & $-900$ & 3.18 & 15.64 & 56.21(26) & 47.21(21) \\
        & $-950$ & 3.24 & 15.93 & 57.44(26) & 48.48(22) \\
        & $-1000$ & 3.28 & 16.13 & 58.30(27) & 49.56(22) \\    
    \end{tabular}
    \caption{Diameters $d_\mathrm{SV}$ of the SV in the low-pressure gas, their equivalents $d_\mathrm{SV}^\mathrm{equ}$ in liquid water with unit density, as well as the mean cluster size $M_1$ and the simulated mean number $N_\mathrm{coll}$ of collected ions for the stated gas pressures $p$ and cathode potentials $U_\mathrm{Cath}$.}
    \label{tab_SV_sizes}
\end{table}

\begin{figure}[thbp]
    \centering
    \includegraphics[width=0.7\linewidth]{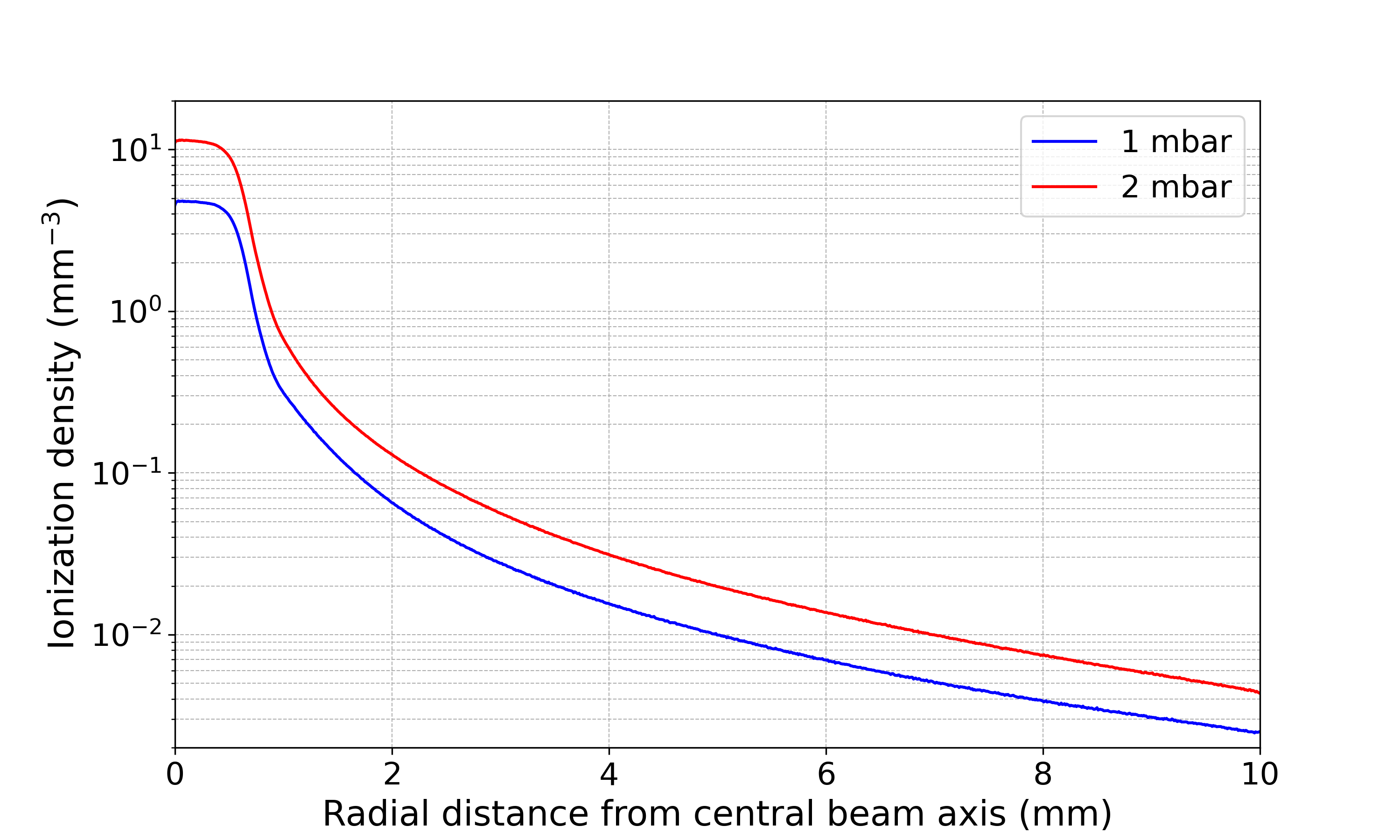}
    \caption{Geant4-DNA simulated ionization density as a function of the radial distance from the central beam axis of the collimated beam of 4.6 MeV alpha particles for the stated propane gas pressures.}
    \label{fig_radial_distribution}
\end{figure}

Another important characteristic of the track structure is the radial distribution of ionizations around the central beam axis of the incident primary particle beam, which directly affects the ion arrival time distribution.
At the pressures applied in this work (1~mbar and 2~mbar), secondary and higher-order delta electrons can travel up to a few centimeters from their point of origin before they thermalize.
Fig.~\ref{fig_radial_distribution} presents the Geant4-DNA simulated ionization density in the drift region above the hole.
Due to the contribution of secondary electrons, the ionization density at radial distances below about 1~mm at 2~mbar is slightly more than double the value at 1~mbar.
This ratio gradually decreases at larger radial distances.
The plateau observed at small radial distances reflects the collimation of the beam.
For radial distances beyond the beam diameter, the ionization density approximately follows an inverse-square dependence on the radial distance.

\subsection{Electric Field, Time-to-Signal and Ion Arrival Time Distributions}\label{results_arrival}
\begin{figure}[thbp]
    \centering
    \includegraphics[width=0.7\linewidth]{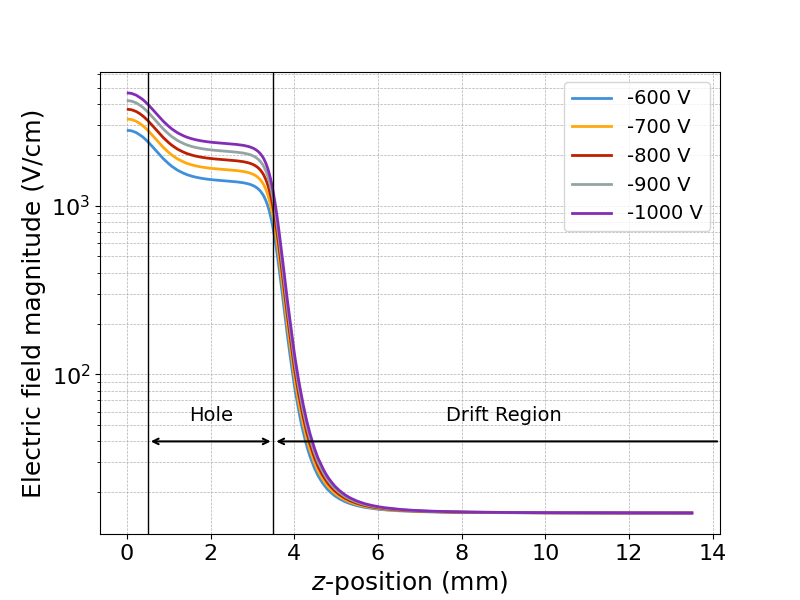}
    \caption{Simulated electric field strength for different cathode potentials along the $z$-axis, which runs centrally through the hole. The central beam axis was at $z=13.5$~mm, the dielectric plate between $z=0.5$~mm and $z=3.5$~mm, and the cathode at $z=0$~mm.}
    \label{fig_efield_comparison}
\end{figure}

Fig.~\ref{fig_efield_comparison} presents the calculated electric field strengths along the $z$-axis of the hole for the applied cathode potentials.
In the drift region, the electric field strength is approximately the same for all cathode potentials with about 15~V/cm.
The contribution of the penetrating field generated by the cathode becomes apparent at about 2~mm above the hole.
The electric field strength increases rapidly in the vicinity of or inside the hole.
The electric field reaches its maximum at the cathode, ranging from 2800~V/cm at a cathode potential of $-600$~V to 4660~V/cm at $-1000$~V.
These values correspond to reduced electric fields $E/N$, where $E$ is the electric field and $N$ the particle density of the gas, of approximately 11500~Td to 19100~Td at 1~mbar and 5750~Td to 9550~Td at 2~mbar propane, respectively, at 24~$^\circ$C.

\begin{figure}[thbp]
    \centering
    \includegraphics[width=0.7\linewidth]{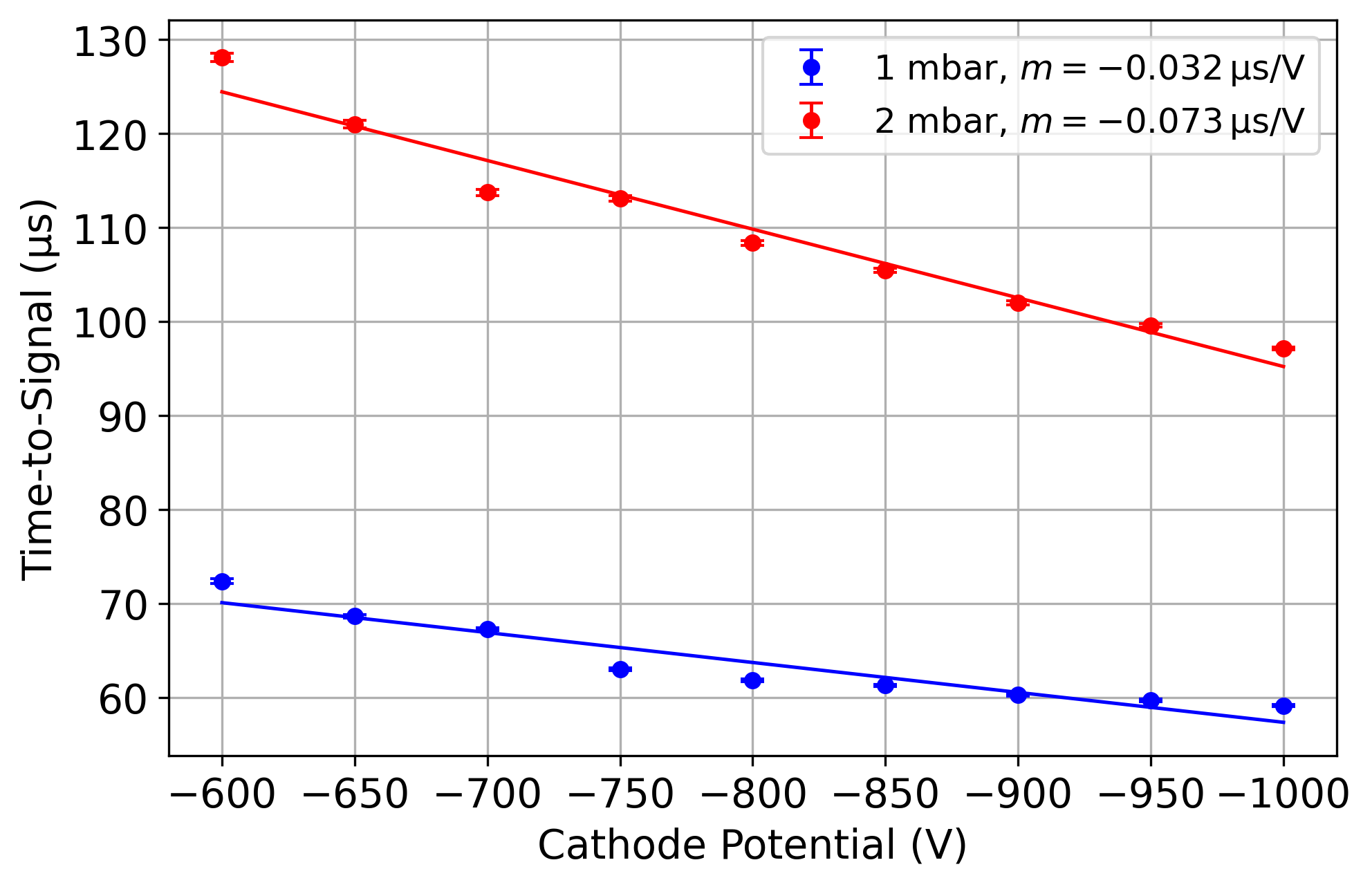}
    \caption{Mean time-to-signal measured as a function of the cathode potential for the stated propane gas pressures. Uncertainty bars are smaller than the symbol size. The solid line represents a linear fit to the data, and $m$ denotes its slope.}
    \label{fig_mean_timetosignal}
\end{figure}

Fig.~\ref{fig_mean_timetosignal} shows the mean time-to-signal measured as a function of the cathode potential.
A decrease in the mean time-to-signal was observed with increasing cathode potential.
The decrease is more pronounced at 2~mbar propane pressure, where the mean time-to-signal drops from approximately 128~\textmu s at $-600$~V to 97~\textmu s at $-1000$~V, compared to 1~mbar, where it decreases from about 72~\textmu s to 59~\textmu s.

The measured time-to-signal distributions are shown in Appendix A, Fig.~\ref{fig_time_to_signal}, (a) and (b).
The distributions are approximately symmetric for both pressures and resemble Gaussian shapes.
At 1~mbar, the standard deviation of the time-to-signal decreases from about 12.0~\textmu s at lower cathode potentials to approximately 10.5~\textmu s at higher cathode potentials. Similarly, at 2~mbar, the standard deviation decreases from about 18.4~\textmu s to 15.4~\textmu s as the cathode potential increases, indicating that the distributions become slightly narrower at higher fields.
The ratio of the standard deviation and the mean of the time-to-signal remains approximately the same for both pressures.

\begin{figure}[thbp]
    \centering
    \includegraphics[width=0.7\linewidth]{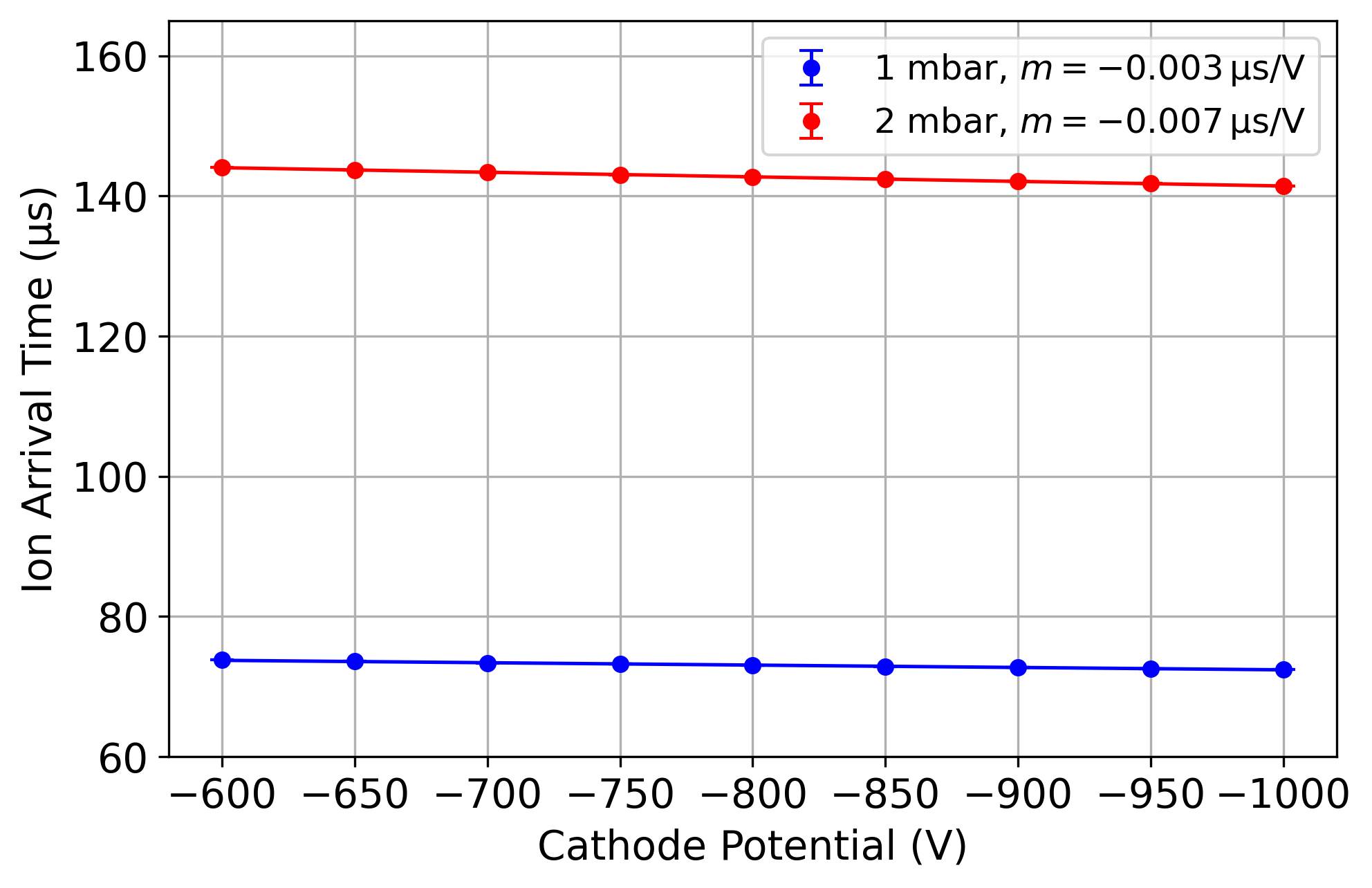}
    \caption{Simulated mean ion arrival times including diffusion for the stated propane gas pressures and cathode potentials. Uncertainty bars are smaller than the symbol size. The solid line represents a linear fit to the data, and $m$ denotes its slope.}
    \label{fig_sim_arrival_time}
\end{figure}

The simulated mean ion arrival times of C$_3$H$_8^+$-ions for the applied detector configurations are shown in Fig.~\ref{fig_sim_arrival_time}.
The mean ion arrival time shows a very weak dependence on the cathode potential, showing a decrease with increasing cathode potential as a direct result of the electric field formation shown above.
At 1~mbar, the mean ion arrival time decreases from about 76~\textmu s to 72~\textmu s with a slope of the linear fit of $-0.003$~\textmu s/V, while at 2~mbar, it decreases from about 148~\textmu s to 141~\textmu s with a slope of $-0.007$~\textmu s/V.
The mean ion arrival time at 2~mbar propane gas pressure is approximately two times the mean ion arrival time at 1~mbar.
Note that the simulated ion arrival time is not representative of the measured time-to-signal.

In Appendix A, Fig.~\ref{fig_sim_arrival_time_spectrum}, (a) and (b), the simulated ion arrival time distributions for the applied detector configurations are presented.
As with the time-to-signal distributions, the width of the Gaussian-like distributions is larger at 2~mbar than at 1~mbar.
The standard deviation of about 10.3~\textmu s at 1~mbar and 19.4~\textmu s at 2~mbar only shows a slight decrease with increasing cathode potential.
Similar to the measured time-to-signal distributions, the ratio of the standard deviation to the mean of the simulated ion arrival time is approximately the same for both pressures.
In this case, we can show that both the standard deviation and the mean increased by approximately the same factor when changing the pressure from 1~mbar to 2~mbar.
This is due to the fact that the spatial spread of the ion starting positions is similar in both cases, as it is primarily determined by the collimated beam of densely ionizing alpha particles.
Note that these distributions represent an ion-by-ion simulation in the static electric field, not including any effects like charging-up of the dielectric or dead time of the detector.

As discussed in the previous section, due to the low pressure of the gas, the radial distance of ionizations from the central beam axis reached the cm range (Fig.~\ref{fig_radial_distribution}).
The initial ion positions in the drift region were distributed accordingly, determining the width of the time-to-signal and the ion arrival time distributions.
Additionally, ion diffusion in the gas results in additional broadening of the distributions.
The repetition of the Garfield++ simulations without diffusion allowed to estimate the contribution of diffusion to the total variance of the ion arrival time.
This contribution was between $\approx 25$~\% at $-600$~V and $\approx 20$~\% at $-1000$~V cathode potential, for both 1~mbar and 2~mbar propane gas pressure, showing a slight decrease with increasing field strength.

All ion drift trajectories simulated using Garfield++ terminated either on the readout electrode, the surrounding copper-covered area, or at the cathode.
No ion impacts on the hole walls were observed in the simulation.

\subsection{Pulse Amplitude and Pulse Area Distributions} \label{res_peak_height_spectra}
\begin{figure}[thbp]
    \centering
    \includegraphics[width=\linewidth]{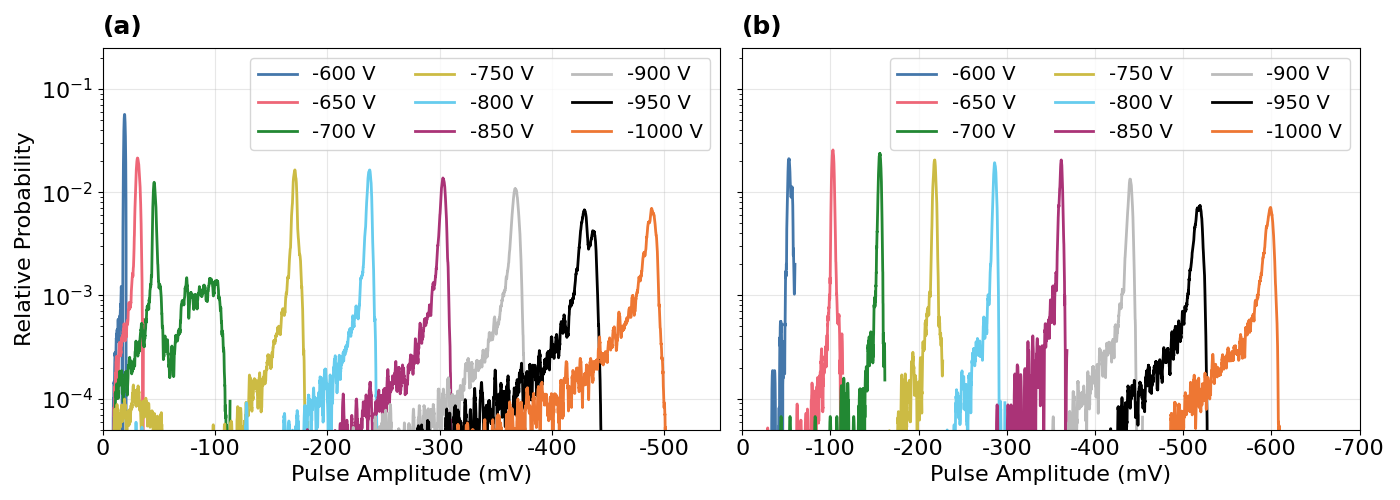}
    \caption{Measured pulse amplitude distributions of the detector signals at 1~mbar (a) and 2~mbar (b) propane gas pressure for the stated cathode potentials.}
    \label{fig_pulse_heights}
\end{figure}

Fig.~\ref{fig_pulse_heights}, (a) and (b), show the measured pulse amplitude distributions at 1~mbar and 2~mbar propane gas pressure.
A clear dependence on both the applied cathode potential and the gas pressure was observed.
For both gas pressures, increasing the cathode potential from $-600$~V to $-1000$~V results in a pronounced shift of the distributions toward higher amplitudes.
All distributions show asymmetric shapes, characterized by a steep fall-off at high amplitudes and an extended tail toward lower amplitudes.
Most detector configurations produce a distinct and well-defined peak corresponding to the most probable pulse amplitude.
At 1~mbar and cathode potentials of $-700$~V, $-750$~V, and $-950$~V, the recorded signals deviate from this behavior: unlike the other configurations, they do not feature a single well-defined peak.
At $-700$~V, a transition is observed in which the initially narrow, closely spaced peaks evolve into broader peaks with larger separations for the same incremental change in cathode potential.
At $-750$~V, additional pulses of lower amplitude appear, although their occurrence is very rare.
At 2~mbar, the peaks are consistently located at higher amplitudes compared to 1~mbar and appear noticeably narrower.
The most probable pulse amplitude increases approximately linearly with the cathode potential at 1~mbar for potentials of $-750\,\rm V$ and above, and at 2~mbar over the entire investigated cathode potential range.

\begin{figure}
    \centering
    \includegraphics[width=\linewidth]{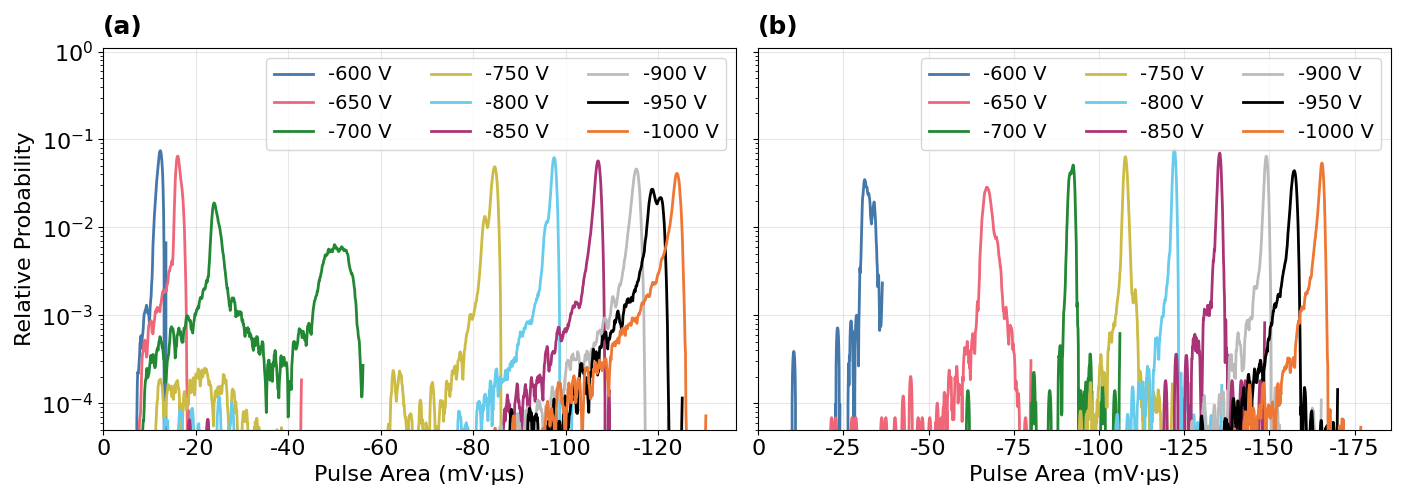}
    \caption{Measured pulse area distributions of the detector signals at 1~mbar (a) and 2~mbar (b) propane gas pressure for the stated cathode potentials.}
    \label{fig_peak_areas}
\end{figure}

Fig.~\ref{fig_peak_areas}, (a) and (b), show the measured pulse area distributions at 1~mbar and 2~mbar propane gas pressure.
As for the pulse amplitude distributions, increasing the cathode potential from $-600$~V to $-1000$~V shifts the distributions toward larger absolute pulse areas, which are generally larger at 2~mbar than at 1~mbar.
The pulse area distributions also exhibit extended tails toward smaller absolute areas and sharp fall-offs toward larger areas.
The complex structures observed at 1~mbar and $-700$~V and $-750$~V are consistent with those in the corresponding pulse amplitude distributions.
Unlike the approximately linear increase in the most probable pulse amplitude over parts of the investigated cathode potential range, the pulse area maxima exhibit a nonlinear dependence on the cathode potential.
At 1~mbar, the maxima are initially closely spaced, with a pronounced transition at $-700$~V, followed by a larger separation at $-750$~V and progressively smaller separations at higher cathode potentials.
At 2~mbar, the spacing between successive maxima gradually decreases with increasing cathode potential.

Overall, the pulse amplitude and pulse area distributions with a single well-defined pulse are remarkably narrow, with the pulses at 2~mbar appearing slightly narrower than those at 1~mbar.
At cathode potentials of $-750\,\rm V$ and above, both the most probable pulse amplitudes and pulse areas are about 75~\% to 81~\% of the values at 2~mbar.

\subsection{Signal Yield}\label{res_signal_yields}
\begin{figure}[thbp]
    \centering
    \includegraphics[width=0.7\linewidth]{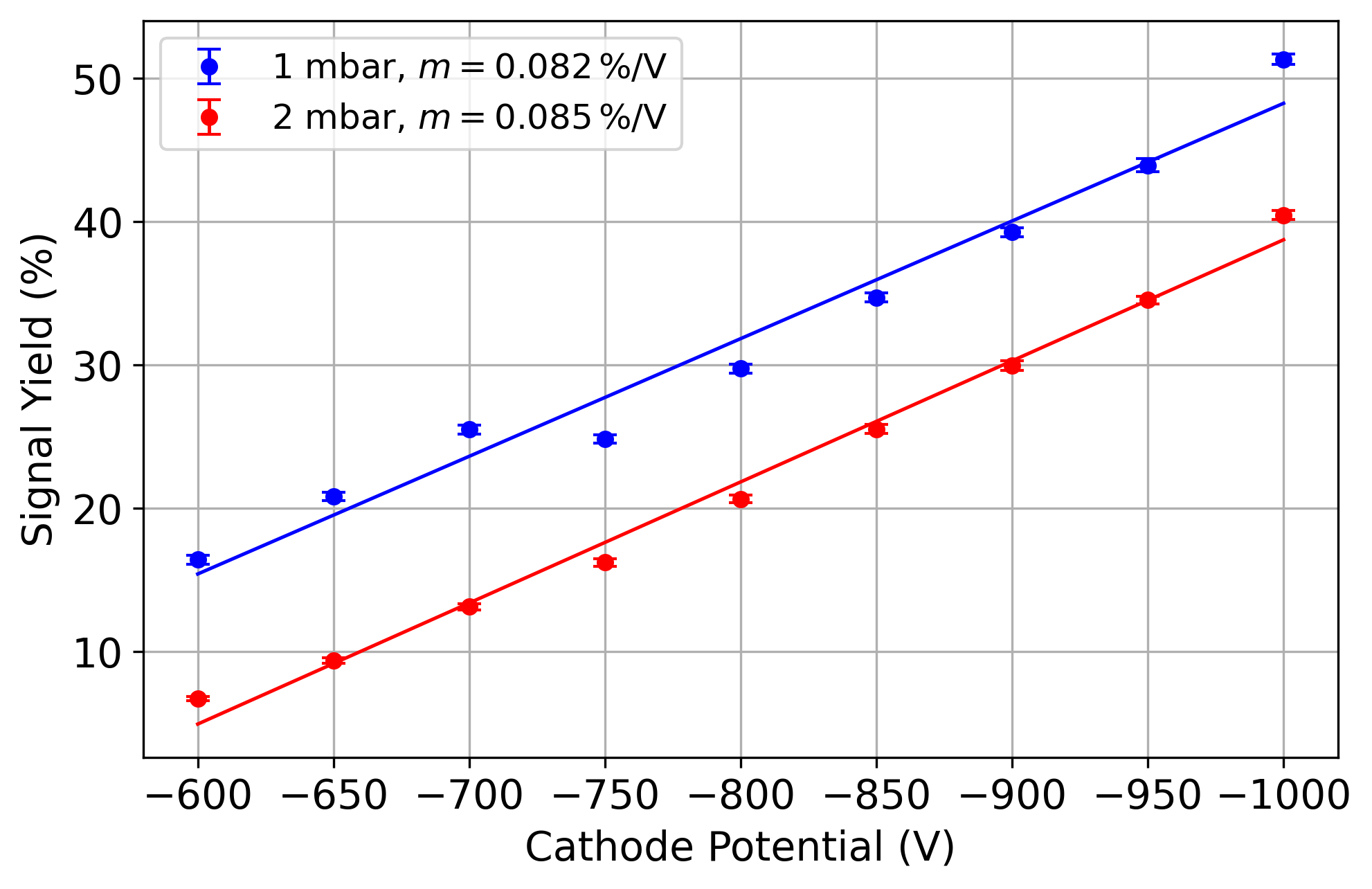}
    \caption{Measured signal yields as a function of the cathode potential for the stated propane gas pressures. Uncertainty bars are smaller than the symbol size. The solid line represents a linear fit to the data, and $m$ denotes its slope.}
    \label{fig_signalyield}
\end{figure}

Fig.~\ref{fig_signalyield} shows the measured signal yield of the detector as a function of the applied cathode potential, for 1~mbar and 2~mbar propane gas pressure, respectively.
The signal yield increases with the applied cathode potential for both 1~mbar and 2~mbar, with nearly identical slopes of the corresponding linear fits of about 0.08~\%/V.
The highest signal yield amounted to about 52~\% at 1~mbar and 41~\% at 2~mbar.

The dark count rate for each configuration was determined with the shutter closed by triggering on the detector signal.
For all configurations, the dark count rates were below 1~min$^{-1}$.
In this work, accepted signals were in the range of $\pm3\sigma$ around the maximum of the Gaussian fit of the time-to-signal distribution (see Sec.~\ref{sec_efficiency}).
This resulted in a reduction of the signal yield by approximately 2~\% for 1~mbar and 3~\% for 2~mbar, which exceeded the expected contribution from dark counts.
Therefore, no subtraction of dark counts was applied.
Under the investigated conditions, no instabilities such as spurious discharges, self-sustained breakdowns, or streamer-like behaviour were observed.
For the applied experimental gas pressures and cathode potentials, the detector response remained stable throughout the measurements.
For cathode potentials exceeding an absolute value of 1000~V, the detector became unstable, exhibiting a high dark count rate of $>10\,\rm s^{-1}$.
Although signals coincident with triggers were still observed, measurements were not continued under these conditions.

\subsection{Signal Formation Probability}
\begin{figure}[thbp]
    \centering
    \includegraphics[width=0.7\linewidth]{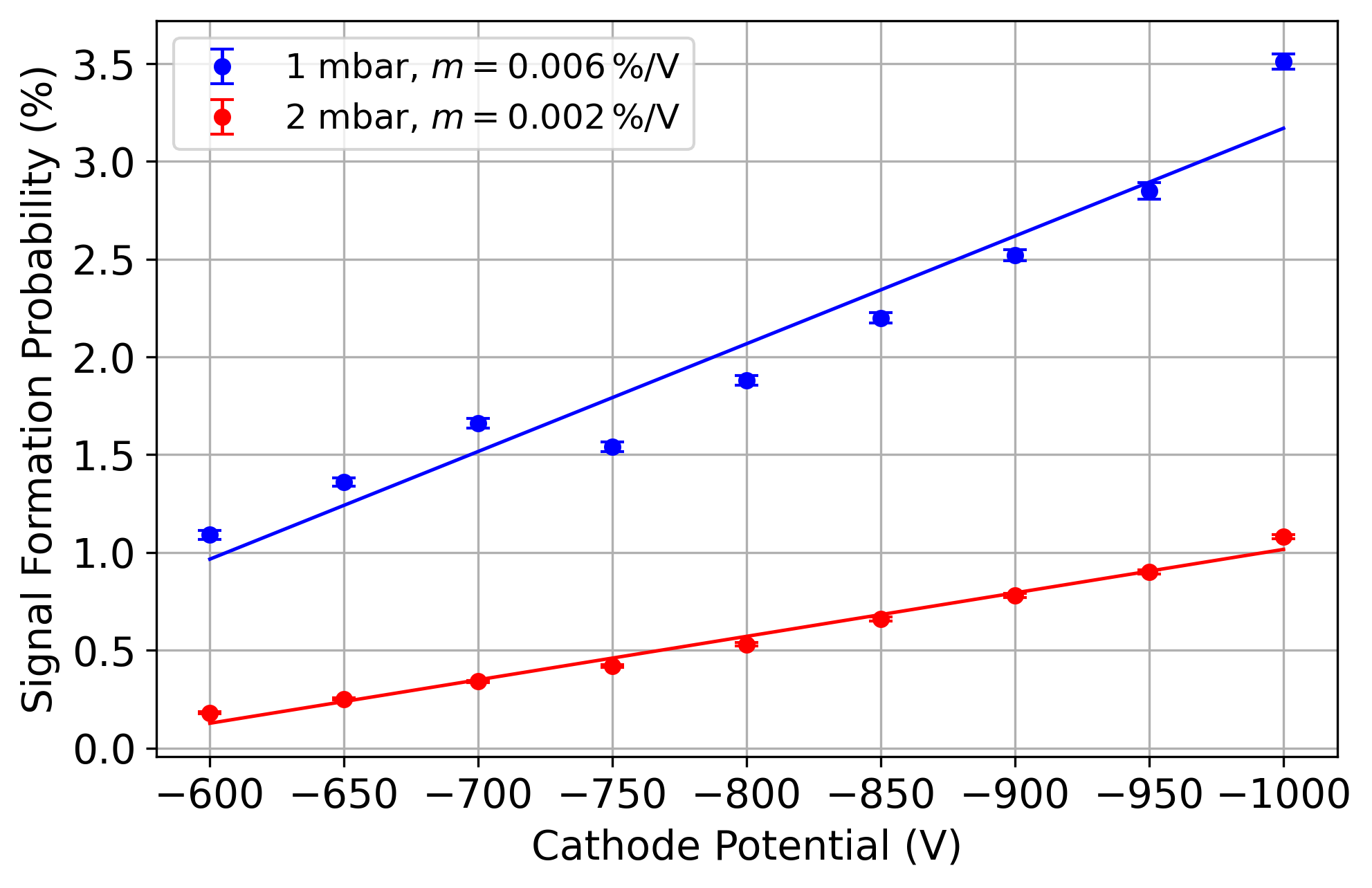}
    \caption{Signal formation probability as a function of the cathode potential for the stated propane gas pressures. The solid line represents a linear fit to the data, and $m$ denotes its slope.}
    \label{fig_ps}
\end{figure}

Fig.~\ref{fig_ps} shows the signal formation probability, calculated with Eq.~\ref{eq_ps}, 
as a function of the cathode potential for 1~mbar and 2~mbar propane gas pressure.
The signal formation probability increases with increasing cathode potential and reaches maximum values of 3.5~\% for 1~mbar and 1.1~\% for 2~mbar.
The slopes of the linear fits to the signal formation probability are 0.06~\%/V for 1~mbar and 0.02~\%/V for 2~mbar, respectively.
Note that the signal formation probability was calculated by combining the measured signal yield and the simulated frequency distribution of the number of collected ions per primary particle, as given in Eq.~\ref{eq_ps}.

\section{Discussion}
\subsection{Signal Yield and Signal Formation Probability}\label{disc_signal_yield}

As shown in Fig.~\ref{fig_signalyield}, the detector prototype signal yield depends strongly on the applied cathode potential; i.e., it increases with the cathode potential to up to 52~\% for 1~mbar and 41~\% for 2~mbar propane gas pressure at $-1000$~V, respectively.
These values are comparable to those of the FIRE \citep{FIRE} and FIRE-V2 \citep{kempf20252} detectors, despite the smaller 3~mm dielectric plate thickness of the present detector.
Higher signal yields may be achievable above $-1000$~V if stable detector operation can be maintained.

At 1 mbar, a deviation from the overall trend toward lower signal yield (Fig.~\ref{fig_signalyield}) and signal formation probabilities (Fig.~\ref{fig_ps}) was observed around $-750\,\rm V$.
Its origin remains unresolved and warrants further investigation.
It may reflect response variation during the measurement campaign.
However, because the estimated signal formation probability remains in the low single-digit percent range, this local deviation does not change the principal result.

While Garfield++ simulations indicate a slight increase in the number of ions collected from the SV with increasing cathode potential (right-most column of Tab.~\ref{tab_SV_sizes}), this effect is too small to account for the pronounced rise in the measured signal yield and signal formation probability.
The stronger electric field inside and above the hole at higher cathode potentials increases the probability that a collected ion initiates a detectable avalanche.
This effect is distinct from the ion collection efficiency calculated in Garfield++, which describes only the transport of ions into the hole and does not include the signal formation processes.

Since the signal yield is determined by the number of ions produced in the SV, it depends on radiation quality.
A systematic investigation of this dependence was beyond the scope of the present study, which focused on characterizing the newly developed prototype and investigating its working principle to the extent possible.
The signal amplitudes and shapes observed here are governed solely by the experimental settings, such as gas type and pressure, cathode potential, and hole geometry, rather than by the radiation quality.

\subsection{Pulse Amplitudes and Pulse Areas}
\label{disc_peak_heights}
The measured pulse amplitude (Fig.~\ref{fig_pulse_heights}) and pulse area (Fig.~\ref{fig_peak_areas}) distributions at different cathode potentials are clearly separated, demonstrating the detector's sensitivity to electric-field variations.
The most probable pulse amplitude increases approximately linearly with cathode potential at 1~mbar for $-750\,\rm V$ and above, and at 2~mbar throughout the investigated range.
In contrast, the most probable pulse area exhibits a nonlinear dependence, suggesting that the pulse shape varies with cathode potential and gas pressure, potentially due to differences in avalanche development associated with the $E/N$ within the hole.
While closely related to detector gain, the pulse area is not a direct measure, as it is also influenced by the readout response.

At 1~mbar, both the pulse amplitude and pulse area distributions are broader and exhibit more pronounced low-amplitude tails, indicating larger statistical fluctuations than at 2~mbar. 
At $-700$ and $-750$~V, pronounced deviations from the otherwise well-defined single-peaked distributions indicate a less uniform detector response.
These coincide with the shifts in signal yield and signal formation probability (Sec.~\ref{disc_signal_yield}), and time-to-signal (Sec.~\ref{disc_temporal_response}).
This may reflect either a measurement variation or inherently unreliable detector behaviour under these operating conditions, although the available data do not allow these possibilities to be distinguished.

The single-peaked distributions are remarkably narrow given the stochastic nature of electron multiplication in gas avalanches, which generally produces substantial fluctuations described by exponential or Polya-type distributions, depending on the electric field configuration and underlying assumptions \citep{Hilke2020}.
The detector's high $E/N$ and low-pressure operation, under conditions associated with Geiger discharges, may lead to behaviour differing from that of conventional detectors.

At 1~mbar, the most probable pulse amplitudes and areas are generally 75~\%--81~\% of those at 2~mbar for cathode potentials of $-750$~V and above.
This modest difference is notable given the factor-of-two pressure difference and its effect on mean free paths and collision rates.
However, reliable simulation of charge avalanches to benchmark the detector response was beyond the scope of this work.
Reliable ionization-coefficient data for the investigated pressures and reduced electric fields are scarce.
Previous propane studies demonstrate a complex pressure dependence of ionization coefficients at high reduced electric fields \citep{bronic_1999}.
Although Bronic and Grosswendt investigated different conditions, their findings at pressures from 20~mbar suggest that nonlinear dependencies of detector response on pressure and reduced electric field may also occur under the present operating conditions.

\subsection{Ion Transport, Temporal Response and Dead Time}
\label{disc_temporal_response}

The simulated ion arrival time(Fig.~\ref{fig_sim_arrival_time}) exhibits almost no dependence on cathode potential, as expected, since the electric field in the drift region remains essentially unchanged and varies only near or within the hole.
Accordingly, cathode potential variations are not expected to affect ion arrival times by more than a few microseconds.
In contrast, the measured time-to-signal (Fig.~\ref{fig_mean_timetosignal}) decreases significantly with increasing cathode potential.
These quantities must be distinguished: the simulated ion arrival time represents the drift time of a $\mathrm{C_3H_8^+}$ ion from its point of origin to the cathode, whereas the measured time-to-signal is the delay between the trigger and detector signal.
Since additional ion species, possible dissociation processes, and the signal formation process itself are not included in the simulations, exact agreement is not expected.
Higher electric fields may increase ion interaction probabilities, potentially leading to earlier signal formation, although no direct evidence currently supports this interpretation.

The standard deviations of the measured time-to-signal (Fig.~\ref{fig_time_to_signal}) and simulated ion arrival-time (Fig.~\ref{fig_sim_arrival_time_spectrum}) distributions exhibit a pressure-dependent behaviour.
At 1~mbar, the measured distribution is broader, whereas at 2~mbar, it is slightly narrower than the simulated distribution.
Additional reaction pathways, such as ion dissociation or the formation of other ion species, could broaden the measured distribution through additional stochastic processes.
While consistent with the observation at 1~mbar, this does not explain the narrower experimental distribution at 2~mbar.
Thus, the effects of additional processes are more complex than a mere increase in statistical variance.
A possible explanation is the selective loss of ions with longer drift paths, originating, for example, in the upper drift region or at larger lateral distances from the hole centre.
Recombination or electric field perturbations could prevent these ions from reaching the hole, thereby narrowing the measured distribution relative to the simulated one.
Such losses are not considered in the simulations, which track ions until they reach a solid surface, irrespective of whether their arrival time exceeds the experimental time window of 200~\textmu s.

These results highlight a limitation of the current Garfield++ simulations of single-ion species, which cannot explain the observed temporal response.
More comprehensive Monte Carlo models incorporating relevant ion-transport and reaction processes, and potentially collective effects and the dependence of signal formation on the local electric field, are needed.
The present findings thus identify key limitations of the existing model and requirements for future studies of the detector's physical response mechanisms.

In addition to the low signal formation efficiency per collected ion, the observation of no more than one signal per trigger indicates a dead time exceeding the 200~\textmu s measurement window.
The most likely cause is a high $RC$ time constant, which determines the cathode recharge time following a voltage breakdown during a Geiger discharge.
The persistence of an ion cloud within the hole, partially screening the cathode electric field, and residual localized charge on the cathode may also contribute.
Consequently, the present detector cannot measure the ICSD, which requires substantially improved ion detection efficiency and a shorter dead time.
Measures to reduce the dead time are currently being considered.

\subsection{Considerations on the Signal Formation Mechanisms}
\label{disc_signal_mechanisms}

Previous work \citep{bashkirov20092} proposed III of gas molecules in the hole as the dominant signal formation mechanism in low-pressure gas-based positive-ion detectors.
However, III cross sections for propane ions are unavailable, leaving its probability and potential dominance under the investigated conditions unknown.
III of neutral gas molecules has nevertheless been observed in previous studies. Pak and Kushner reported III in helium at pressures of approximately 0.13~mbar to a few mbar, using cathode potentials of several tens of kilovolts and electrode separations of a few millimeters \citep{pak1992}. Although these findings demonstrate the feasibility of III under specific conditions, particularly at higher ion energies than in the present setup, they do not establish its contribution to the observed detector response.

In general, both the kinetic and potential energy of incoming ions can induce IISEE from surfaces such as the hole walls or cathode. This mechanism has not been discussed in detail in previous publications on compact nanodosimeters based on the same working principle.

Garfield++ simulations indicate that ion drift trajectories are concentrated in the centre of the hole, practically excluding a significant contribution from IISEE at the hole walls for the investigated geometry.

Considering the collision kinetics of drifting ions, their kinetic energy upon reaching the cathode remains below the keV range, whereas kinetic energies in the keV range are typically required for kinetic IISEE \citep{Varga1992}. This mechanism can therefore also be excluded.

IISEE at the cathode through potential-energy transfer remains possible if the ion potential energy exceeds approximately twice the cathode work function \citep{Varga1992}. For $\mathrm{C_3H_8^+}$ ions, with an ionization potential of 11.1~eV, this requires a cathode energy barrier below approximately 5.55~eV. Surface deposits and impurities may affect the IISEE yield, although they generally tend to reduce electron yields \citep{Arazi2018}. Only a few studies report IISEE yields for insulators, including glass \citep{Elsbergen2000} and other materials \citep{Kim2000,CERN_Tech_Note}, with values reaching approximately 50~\% for MgO in the relevant energy range. However, these measurements did not involve propane ions, and available data generally indicate lower yields for molecular ions and ions in gases other than noble gases \citep{Vance1968,Vance1968_2}. Notably, the ionization energy of $\rm C_3H_8$ exceeds the cathode work function, making IISEE energetically more favourable than III.

IISEE driven by potential-energy transfer is generally expected to depend only weakly on ion kinetic energy \citep{Varga1992}.
Conversely, the measured signal yield (Fig.~\ref{fig_signalyield}) and signal formation probability (Fig.~\ref{fig_ps}) increase markedly with cathode potential and are higher at 1~mbar than at 2~mbar, where $E/N$ is twice as high.
This is notable because 2~mbar is closer to the Paschen minimum for propane, which would generally favour avalanche development, and the simulations predict more collected ions at this pressure (Tab.~\ref{tab_SV_sizes}). If IISEE dominated signal formation, the larger number of collected ions might therefore be expected to increase the signal yield.
The lower signal yield and signal formation probability at 2~mbar may instead be consistent with a contribution from III, as more frequent collisions at higher pressure reduce the kinetic energy gained by ions between collisions. However, no direct evidence for this mechanism is provided by the present data. Other unidentified effects inhibiting avalanche development at higher pressures or lower cathode potentials, as well as non-equilibrium effects, particularly at 1~mbar, may also contribute.


Vasi et al.~\citep{vasi2016} reported increasing detector efficiency with increasing dielectric plate thickness at constant $E/N$, using a detector with similar conditions, including low-pressure propane, a glass cathode and alpha-particle irradiation. This behaviour would be consistent with III, since a thicker plate increases the probability of ion-impact ionization but not that of IISEE at the cathode. This provides an additional indication of a possible III contribution, although differences in detector geometry and operating conditions must be considered.

Overall, the present measurements do not allow the contributions of III and IISEE to be quantified or distinguished unambiguously. Although III may contribute to the detector response, no direct evidence for its occurrence is available. IISEE at the cathode, not previously considered in the relevant literature on positive-ion detectors, must therefore also be considered as a potential signal formation mechanism. Dedicated measurements and simulations are needed to establish the underlying physical processes.

\subsection{Ongoing and Future Studies}
\label{disc_future}
 A long-term goal of compact low-pressure gas detectors based on positive-ion detection is to develop nanodosimeters capable of measuring ICSDs.
The feasibility of ICSD measurements depends on ion-counting capability, governed by the signal formation probability per ion and detector dead time. Increasing the former while reducing the latter is therefore essential for future detector development.
Potential approaches include optimizing the dielectric and cathode materials, detector geometry, working gas and pressure, as well as implementing layered dielectric-electrode structures or resistive coatings to tailor the electric field inside the hole.


Multi-hole amplification structures with independently operated holes could further complement these developments towards ICSD measurements. Sufficient overlap of the SVs associated with individual holes could enable counting individual ions originating within a common nanometer-equivalent SV. This approach could employ independently biased holes with sub-millimeter resistive cathodes. A first step was demonstrated in~\citep{merza2026}, using multi-hole amplification structures with a common cathode.

For quantitative ICSD measurements, detection efficiencies below 100~\% require correction for missed ions, e.g., through binomial reconstruction assuming independent detection with known efficiency. Lower efficiencies increase reconstruction uncertainties, particularly for higher moments, and may require large data sets and long acquisition times. Improving and accurately determining the signal-formation probability is therefore essential for reliable ICSD reconstruction of future prototypes.

Dedicated measurements of IISEE yields for detector-relevant cathode materials and ion species are needed to quantify its contribution. Further experiments should investigate the influence of cathode material and working gas on signal yield. Cathode heating or coatings with materials of known IISEE yield could provide additional means of assessing surface-induced electron emission. Cathodes with adjustable resistance may also help minimize dead time. Accurate simulations of ion kinetic energies along their drift trajectories are required to assess the plausibility of the proposed signal formation mechanisms. Although measuring III cross sections and modelling their contribution under relevant operating conditions are more challenging, the contribution of III to signal formation must ultimately be determined to model detector response.



\section{Conclusion}
A low-pressure gas-based positive-ion detector was developed and characterized experimentally and computationally.
A key achievement was the quantification of the per-ion signal formation probability, which was found to be only a few percent.
The detector response was further critically assessed with respect to the potential contributions of III and IISEE.
However, the available evidence does not allow an unambiguous identification of the dominant signal formation mechanism, requiring further experimental investigations and computational studies.

\section{Author statement}
\textbf{Victor Merza:} Writing - Original Draft, Methodology, Conceptualization, Software.
\textbf{Aleksandr Bancer:} Conceptualization, Resources.
\textbf{Vladimir Bashkirov:} Conceptualization, Methodology.
\textbf{Ana Belchior:} Supervision.
\textbf{Beata Brzozowska:} Writing - Review \& Editing, Methodology, Project administration.
\textbf{João F. Canhoto:} Writing - Review \& Editing.
\textbf{Piotr Gasik:} Writing - Review \& Editing, Methodology.
\textbf{Jaroslaw Grzyb:} Methodology, Resources.
\textbf{Khaled Katmeh:} Writing - Review \& Editing.
\textbf{Marcin Pietrzak:} Writing - Review \& Editing, Methodology, Software.
\textbf{Antoni Ruciński:} Supervision, Project administration.
\textbf{Reinhard Schulte:} Writing - Review \& Editing, Conceptualization, Supervision.

\section{Data statement}
All data can be provided by the corresponding author upon request.

\section{Acknowledgements}
This work was partially funded by the Fundação para a Ciência e a Tecnologia (FCT) through the research grants PRT/BD/153748/2021, PRT/BD/151544/2021, and PRT/BD/153750/2021.
Further, financial support was received from the National Science Centre, Poland, under the grant number UMO-024/06/Y/ST2/00196.
The authors gratefully acknowledge the support of Ingo Deppner (GSI), who kindly provided the low-resistivity glass used in this work.

\appendix

\section{Additional Figures}
The figures presented below support the results and discussion presented in the main text.

\begin{figure}[htpb]
    \centering
    \includegraphics[width=\textwidth]{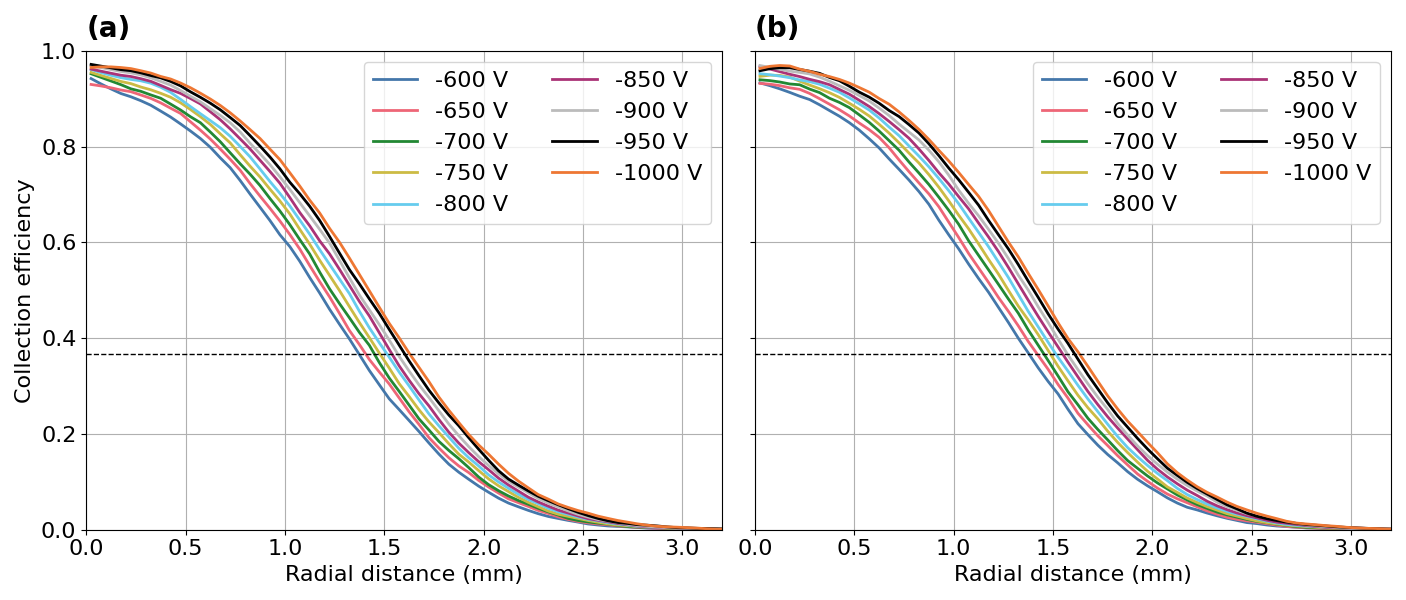}
    \caption{Simulated radial ion collection efficiencies for the stated cathode potentials in 1~mbar (a) and 2~mbar (b) propane. The dashed line corresponds to a collection efficiency of 1/$e$ and determines the nominal size of the SV.}
    \label{fig_collection_efficiency}
\end{figure}

\begin{figure}[htpb]
    \centering
    \includegraphics[width=\textwidth]{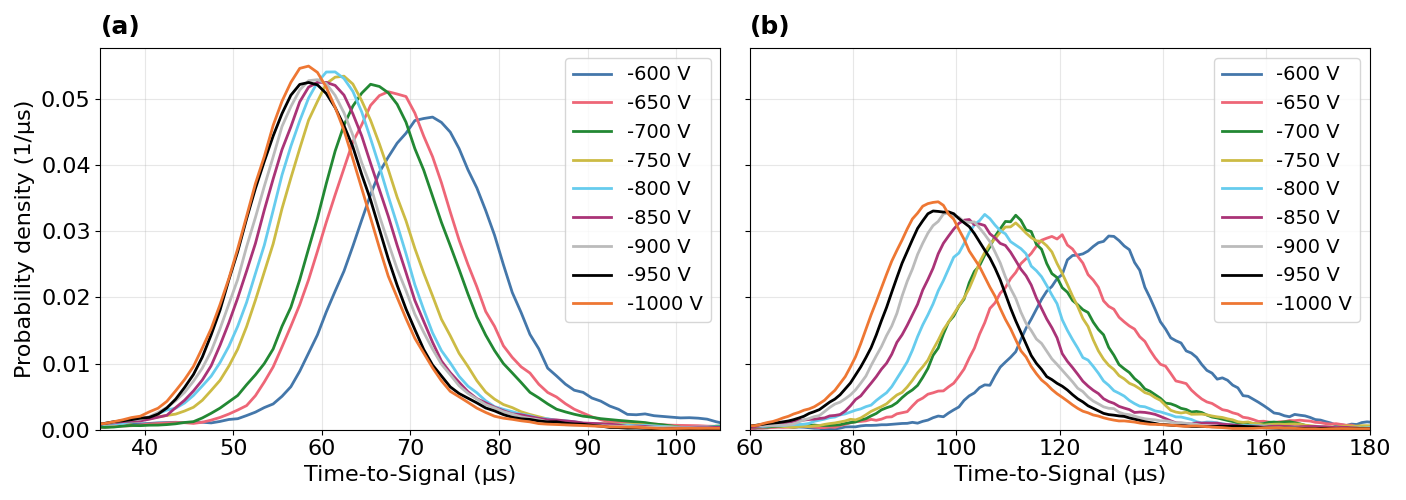}
    \caption{Measured time-to-signal distributions in 1~mbar (a) and 2~mbar (b) propane for the stated cathode potentials.}
    \label{fig_time_to_signal}
\end{figure}

\begin{figure}[htpb]
    \centering
    \includegraphics[width=\textwidth]{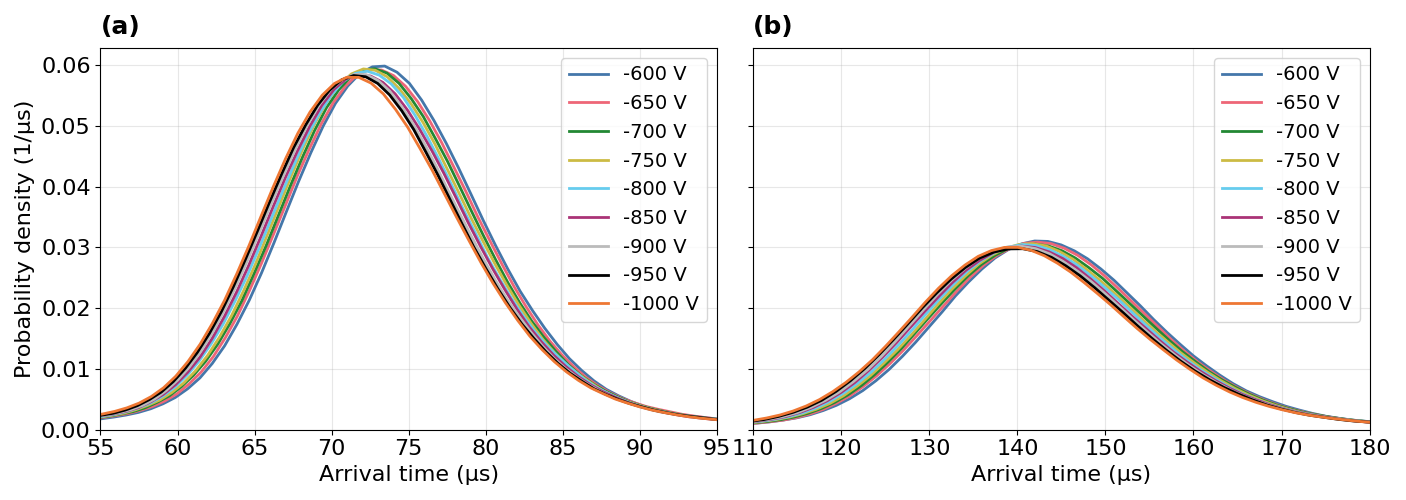}
    \caption{Simulated ion arrival time distributions including diffusion for the stated cathode potentials at 1~mbar (a) and 2~mbar (b) propane gas pressure.}
    \label{fig_sim_arrival_time_spectrum}
\end{figure}

\FloatBarrier
\section{Experimental and Simulation Uncertainties}\label{app_uncertainties}
The stated uncertainties in the Results are equal to the standard uncertainties of the quantities as discussed below.
A coverage factor of $k=1$ was used, which corresponds to a confidence interval of 68 \% \citep{GUM}.

\subsection{Experimental Signal Yield}\label{corr_sign_yield}
The signal yield $\eta$ was calculated as stated in Eq.~\ref{equ_efficiency}.
$N$ is the number of primary particles registered by the trigger, $t$ is the time window opened after each trigger in which signals were recorded, and $N_\mathrm{coinc}$ is the number of signals recorded within the arrival time window $t_\mathrm{arr}$.
Of note, during our experiments, only one or no detector signal was recorded within the time window $t$ following a trigger.
The partial derivative of $\eta$ with respect to $N_\mathrm{coinc}$ is
\begin{equation}
\frac{\mathrm{d} \eta}{\mathrm{d} N_\mathrm{coinc}}=\frac{1}{N}
\end{equation}
Assuming a binomial distribution of $N_\mathrm{coinc}$, the variance of $N_\mathrm{coinc}$ is given by
\begin{equation}
\sigma^2_{N_\mathrm{coinc}}=N\cdot\eta\cdot(1-\eta)
\end{equation}
Applying the Gaussian error propagation, the standard uncertainty $u_\eta$ of $\eta$ thus resulted in
\begin{equation}
u_\eta=\sqrt{\left(\frac{\mathrm{d} \eta}{\mathrm{d} N_\mathrm{coinc}}\right)^2 \cdot \sigma_{N_\mathrm{coinc}}^2} = \sqrt{\frac{\eta\cdot(1-\eta)}{N}}
\end{equation}

\subsection{Simulated ICSD}
The standard uncertainty $u_{M_1}$ of $M_1$ of the presented ICSD is given by
\begin{equation}
u_{M_1}=\frac{\sigma_\nu}{\sqrt{N}}
\end{equation}
where $\sigma_\nu$ is the standard deviation of the cluster size $\nu$ and $N$ is the total number of starting primary particles that produced that ICSD.
The standard uncertainty $u_{p_\nu}$ of the individual probability $p_\nu$ of the cluster size $\nu$ is given by
\begin{equation}
    u_{p_\nu}=\sqrt{\frac{{p_\nu \cdot (1-p_\nu)}}{N}}
\end{equation}

\subsection{Estimated Signal formation Probability}
Since Eq.~\ref{eq_ps} is a nonlinear relation with respect to the signal formation probability $p_\mathrm{s}$, implicit differentiation is required to obtain the standard uncertainty $u_{p_\mathrm{s}}$ of $p_\mathrm{s}$.
We define the function $f(p_\mathrm{s}(\eta,n_i))$:
\begin{equation}
    f(p_\mathrm{s}(\eta,n_i)) = \frac{1}{N}\cdot\sum_{i=1}^N (1-p_\mathrm{s})^{n_i} - (1-\eta) = 0
\end{equation}
In the following, we leave the functional dependence of $f(p_\mathrm{s}(\eta,n_i))$ implicit for better readability.
Further, we use the short-hand notation $S=\frac{1}{N}\cdot\sum_{i=1}^N n_i\cdot(1-p_\mathrm{s})^{n_i-1}$.
The partial derivatives of $p_\mathrm{s}$ with respect to $\eta$ and $n_i$ result in:
\begin{equation}
    \frac{\partial p_\mathrm{s}}{\partial \eta} = -\frac{\partial f/\partial\eta}{\partial f/\partial p_\mathrm{s}} = \frac{1}{S}
\end{equation}
and, approximating $n_i$ as a continuous function,
\begin{equation}
    \frac{\partial p_\mathrm{s}}{\partial n_i} = -\frac{\partial f/\partial n_i}{\partial f/\partial p_\mathrm{s}} = \frac{(1-p_\mathrm{s})^{n_i}\cdot\mathrm{ln}(1-p_\mathrm{s})}{N\cdot S}
\end{equation}
According to the Gaussian error propagation law, the combined standard uncertainty $u_{p_\mathrm{s}}$ is given by:
\begin{equation}
    u_{p_\mathrm{s}} = \sqrt{ \left( \frac{\partial p_\mathrm{s}}{\partial \eta} \right)^2\cdot u_\eta^2 + \sum_{i=1}^N \left( \frac{\partial p_\mathrm{s}}{\partial n_i} \right)^2\cdot u_{n_i}^2}
\end{equation}
The standard uncertainty $u_{n_i}$ of the simulated number $n_i$ of collected ions for each primary particle, which was obtained from the Garfield++ simulation, is given by $u_{n_i} = \sqrt{n_i}$.
The standard uncertainty $u_{n_\eta}$ was determined in \ref{corr_sign_yield}.
Finally, $u_{p_\mathrm{s}}$ results in
\begin{equation}
    u_{p_\mathrm{s}} = \sqrt{ \frac{1}{S^2} \cdot u_\eta^2 + \sum_{i=1}^N   \frac{(1-p_\mathrm{s})^{2n_i}\cdot(\mathrm{ln}(1-p_\mathrm{s}))^2}{N^2\cdot S^2} \cdot n_i}
\end{equation}

\subsection{Other Experimental and Simulated Quantities}
For the additional experimental and simulated quantities determined in this work, e.g., the measured time-to-signal or the simulated number of ions collected in the hole, the stated uncertainty represents the standard uncertainty $u$ of the mean value of the quantity, which is given by
\begin{equation}
    u = \frac{\sigma}{\sqrt{N}}
\end{equation}
where $\sigma$ is the standard deviation and $N$ is the number of measured or simulated samples.

\bibliographystyle{elsarticle-harv}
\bibliography{paper.bib}

@article{rucinski,
doi = {10.1088/1361-6560/ac35f1},
url = {https://dx.doi.org/10.1088/1361-6560/ac35f1},
year = {2021},
month = {dec},
publisher = {IOP Publishing},
volume = {66},
number = {24},
pages = {24TR01},
author = {Antoni Rucinski and Anna Biernacka and Reinhard Schulte},
title = {Applications of nanodosimetry in particle therapy planning and beyond},
journal = {Physics in Medicine \& Biology},
}

@article{Faddegon,
doi = {10.1088/1361-6560/acea16},
url = {https://dx.doi.org/10.1088/1361-6560/acea16},
year = {2023},
month = {aug},
publisher = {IOP Publishing},
volume = {68},
number = {17},
pages = {175013},
author = {Bruce Faddegon and Eleanor A Blakely and Lucas Burigo and Yair Censor and Ivana Dokic and Naoki Dom\'inguez Kondo and Ramon Ortiz and Jos\'e Ramos M\'endez and Antoni Rucinski and Keith Schubert and Niklas Wahl and Reinhard Schulte},
title = {Ionization detail parameters and cluster dose: a mathematical model for selection of nanodosimetric quantities for use in treatment planning in charged particle radiotherapy},
journal = {Physics in Medicine \& Biology}
}

@article{garty2002,
title = {The performance of a novel ion-counting nanodosimeter},
journal = {Nuclear Instruments and Methods in Physics Research Section A: Accelerators, Spectrometers, Detectors and Associated Equipment},
volume = {492},
number = {1},
pages = {212-235},
year = {2002},
issn = {0168-9002},
doi = {https://doi.org/10.1016/S0168-9002(02)01278-0},
url = {https://www.sciencedirect.com/science/article/pii/S0168900202012780},
author = {G Garty and S Shchemelinin and A Breskin and R Chechik and G Assaf and I Orion and V Bashkirov and R Schulte and B Grosswendt}
}

@article{hilgers2015,
title = {Secondary ionisations in a wall-less ion-counting nanodosimeter: quantitative analysis and the effect on the comparison of measured and simulated track structure parameters in nanometric volumes},
journal = {The European Physical Journal D},
volume = {69},
number = {239},
year = {2015},
doi = {10.1140/epjd/e2015-60176-6},
url = {https://doi.org/10.1140/epjd/e2015-60176-6},
author = {G Hilgers and M U Bug and E Gargioni and H Rabus}
}

@article{hilgers2017,
doi = {10.1088/1361-6560/aa86e8},
url = {https://doi.org/10.1088/1361-6560/aa86e8},
year = {2017},
month = {sep},
publisher = {IOP Publishing},
volume = {62},
number = {19},
pages = {7569},
author = {Hilgers, G and Bug, M U and Rabus, H},
title = {Measurement of track structure parameters of low and medium energy helium and carbon ions in nanometric volumes},
journal = {Physics in Medicine \& Biology}
}

@article{hilgers2019,
doi = {10.1088/1748-0221/14/07/P07012},
url = {https://dx.doi.org/10.1088/1748-0221/14/07/P07012},
year = {2019},
month = {jul},
publisher = {},
volume = {14},
number = {07},
pages = {P07012},
author = {G. Hilgers and H. Rabus},
title = {Reducing the background of secondary ions in an ion-counting nanodosimeter},
journal = {Journal of Instrumentation}
}

@article{hilgers2022,
title = {Characterisation of the PTB ion counter nanodosimeter's target volume and its equivalent size in terms of liquid H2O},
journal = {Radiation Physics and Chemistry},
volume = {191},
pages = {109862},
year = {2022},
issn = {0969-806X},
doi = {https://doi.org/10.1016/j.radphyschem.2021.109862},
url = {https://www.sciencedirect.com/science/article/pii/S0969806X21005120},
author = {G Hilgers and T Braunroth and H Rabus}
}

@article{denardo2002,
title = {Ionization-cluster distributions of α-particles in nanometric volumes of propane: measurement and calculation},
journal = {Radiation and Environmental Biophysics},
volume = {41},
pages = {235-256},
year = {2002},
issn = {1432-2099},
doi = {10.1007/s00411-002-0171-6},
url = {https://doi.org/10.1007/s00411-002-0171-6},
author = {L {De Nardo} and P Colautti and V Conte and W Baek and B Grosswendt and G Tornielli}
}

@article{pszona2000,
title = {A new method for measuring ion clusters produced by charged particles in nanometre track sections of DNA size},
journal = {Nuclear Instruments and Methods in Physics Research Section A: Accelerators, Spectrometers, Detectors and Associated Equipment},
volume = {447},
number = {3},
pages = {601-607},
year = {2000},
issn = {0168-9002},
doi = {https://doi.org/10.1016/S0168-9002(99)01191-2},
url = {https://www.sciencedirect.com/science/article/pii/S0168900299011912},
author = {S Pszona and J Kula and S Marjanska}
}

@article{casiraghi2015,
    author = {Casiraghi, M. and Bashkirov, V. A. and Hurley, R. F. and Schulte, R. W.},
    title = "{Characterisation of a track structure imaging detector}",
    journal = {Radiation Protection Dosimetry},
    volume = {166},
    number = {1-4},
    pages = {223-227},
    year = {2015},
    month = {04},
    issn = {0144-8420},
    doi = {10.1093/rpd/ncv139},
    url = {https://doi.org/10.1093/rpd/ncv139},
    eprint = {https://academic.oup.com/rpd/article-pdf/166/1-4/223/4565335/ncv139.pdf},
}

@article{FIRE,
title = {FIRE: A compact nanodosimeter detector based on ion amplification in gas},
journal = {Nuclear Instruments and Methods in Physics Research Section A: Accelerators, Spectrometers, Detectors and Associated Equipment},
volume = {999},
pages = {165116},
year = {2021},
issn = {0168-9002},
doi = {https://doi.org/10.1016/j.nima.2021.165116},
url = {https://www.sciencedirect.com/science/article/pii/S0168900221001005},
author = {Fabiano Vasi and Irina Kempf and J\"urgen Besserer and Uwe Schneider}
}

@article{monte_carlo_model,
title = {Monte Carlo model for ion mobility and diffusion for characteristic electric fields in nanodosimetry},
journal = {Zeitschrift f\"ur Medizinische Physik},
volume = {34},
number = {1},
pages = {140-152},
year = {2024},
note = {Special Issue: Space Radiation Research},
issn = {0939-3889},
doi = {https://doi.org/10.1016/j.zemedi.2022.12.006},
url = {https://www.sciencedirect.com/science/article/pii/S0939388922001398},
author = {Irina Kempf and Uwe Schneider}
}

@article{grosswendt1,
    author = {Grosswendt, B.},
    title = "{Recent advances of nanodosimetry}",
    journal = {Radiation Protection Dosimetry},
    volume = {110},
    number = {1-4},
    pages = {789-799},
    year = {2004},
    month = {08},
    issn = {0144-8420},
    doi = {10.1093/rpd/nch171},
    url = {https://doi.org/10.1093/rpd/nch171},
    eprint = {https://academic.oup.com/rpd/article-pdf/110/1-4/789/4529014/nch171.pdf},
}

@article{grosswendt2,
    author = {Grosswendt, B. and De Nardo, L. and Colautti, P. and Pszona, S. and Conte, V. and Tornielli, G.},
    title = "{Experimental equivalent cluster-size distributions in nanometric volumes of liquid water}",
    journal = {Radiation Protection Dosimetry},
    volume = {110},
    number = {1-4},
    pages = {851-857},
    year = {2004},
    month = {08},
    issn = {0144-8420},
    doi = {10.1093/rpd/nch203},
    url = {https://doi.org/10.1093/rpd/nch203},
    eprint = {https://academic.oup.com/rpd/article-pdf/110/1-4/851/4529938/nch203.pdf},
}

@phdthesis{bantsar,
    author = {Bantsar, Aliaksandr},
    title = "{Ionization Cluster Size Distributions Created by Low Energy Electrons and Alpha Particles in Nanometric Track Segment in Gases}",
    eprint = {1207.6893},
    archivePrefix = {arXiv},
    primaryClass = {physics.atom-ph},
    school = {Soltan Inst., Swierk},
    year = {2010}
}

@manual{elmerfem,
    author = {M. Malinen and P. R\r{a}back},
    title = {Elmer finite element solver for multiphysics and multiscale problems. In book: Multiscale Modelling Methods for Applications in Material Science},
    year = {2013},
pages = {101-113},
chapter = {Elmer finite element solver for multiphysics and multiscale problems},
}

@manual{garfieldpp,
    author = {H. Schindler},
    title = {Garfield++ User Guide Version 2025.1},
    year = {2025},
    url = {https://garfieldpp.web.cern.ch/garfieldpp/},
}

@ARTICLE{bashkirov20091,
  author={Bashkirov, V. and Schulte, R. and Wroe, A. and Sadrozinski, H. and Gargioni, E. and Grosswendt, B.},
  journal={IEEE Transactions on Nuclear Science}, 
  title={Experimental Validation of Track Structure Models}, 
  year={2009},
  volume={56},
  number={5},
  pages={2859-2863},
  doi={10.1109/TNS.2009.2029574}}

@INPROCEEDINGS{bashkirov20092,
  author={Bashkirov, V. A. and Hurley, R. F. and Schulte, R. W.},
  booktitle={2009 IEEE Nuclear Science Symposium Conference Record (NSS/MIC)}, 
  title={A novel detector for 2D ion detection in low-pressure gas and its applications}, 
  year={2009},
  volume={},
  number={},
  pages={694-698},
  doi={10.1109/NSSMIC.2009.5402061}}

@article{casiraghi2014,
	Author = {Casiraghi, Margherita and Bashkirov, Vladimir and Hurley, Ford and Schulte, Reinhard},
	Da = {2014/05/07},
	Doi = {10.1140/epjd/e2014-40841-0},
	Id = {Casiraghi2014},
	Isbn = {1434-6079},
	Journal = {The European Physical Journal D},
	Number = {5},
	Pages = {111},
	Title = {A novel approach to study radiation track structure with nanometer-equivalent resolution},
	Ty = {JOUR},
	Url = {https://doi.org/10.1140/epjd/e2014-40841-0},
	Volume = {68},
	Year = {2014}}

@article{vasi2016,
doi = {10.1088/1748-0221/11/09/C09021},
url = {https://dx.doi.org/10.1088/1748-0221/11/09/C09021},
year = {2016},
month = {sep},
publisher = {},
volume = {11},
number = {09},
pages = {C09021},
author = {F. Vasi and M. Casiraghi and V. Bashkirov and U. Giesen and R.W. Schulte},
title = {Development of a single ion detector for radiation track structure studies},
journal = {Journal of Instrumentation}
}

@article{conte2017,
    author = {Conte, V and Selva, A and Colautti, P and Hilgers, G and Rabus, H and Bantsar, A and Pietrzak, M and Pszona, S},
    title = "{NANODOSIMETRY: TOWARDS A NEW CONCEPT OF RADIATION QUALITY}",
    journal = {Radiation Protection Dosimetry},
    volume = {180},
    number = {1-4},
    pages = {150-156},
    year = {2017},
    month = {09},
    issn = {0144-8420},
    doi = {10.1093/rpd/ncx175},
    url = {https://doi.org/10.1093/rpd/ncx175},
    eprint = {https://academic.oup.com/rpd/article-pdf/180/1-4/150/25409814/ncx175.pdf},
}

@article{G4DNA1,
author = {Tran, Hoang Ngoc and Archer, Jay and Baldacchino, Gérard and Brown, Jeremy M. C. and Chappuis, Flore and Cirrone, Giuseppe Antonio Pablo and Desorgher, Laurent and Dominguez, Naoki and Fattori, Serena and Guatelli, Susanna and Ivantchenko, Vladimir and Méndez, JosÃ©-Ramos and Nieminen, Petteri and Perrot, Yann and Sakata, Dousatsu and Santin, Giovanni and Shin, Wook-Geun and Villagrasa, Carmen and Zein, Sara and Incerti, Sebastien},
title = {Review of chemical models and applications in Geant4-DNA: Report from the ESA BioRad III Project},
journal = {Medical Physics},
volume = {51},
number = {9},
pages = {5873-5889},
doi = {https://doi.org/10.1002/mp.17256},
url = {https://aapm.onlinelibrary.wiley.com/doi/abs/10.1002/mp.17256},
eprint = {https://aapm.onlinelibrary.wiley.com/doi/pdf/10.1002/mp.17256},
year = {2024}
}

@article{G4DNA2,
author = {Incerti, S. and Kyriakou, I. and Bernal, M. A. and Bordage, M. C. and Francis, Z. and Guatelli, S. and Ivanchenko, V. and Karamitros, M. and Lampe, N. and Lee, S. B. and Meylan, S. and Min, C. H. and Shin, W. G. and Nieminen, P. and Sakata, D. and Tang, N. and Villagrasa, C. and Tran, H. N. and Brown, J. M. C.},
title = {Geant4-DNA example applications for track structure simulations in liquid water: A report from the Geant4-DNA Project},
journal = {Medical Physics},
volume = {45},
number = {8},
pages = {e722-e739},
doi = {https://doi.org/10.1002/mp.13048},
url = {https://aapm.onlinelibrary.wiley.com/doi/abs/10.1002/mp.13048},
eprint = {https://aapm.onlinelibrary.wiley.com/doi/pdf/10.1002/mp.13048},
year = {2018}
}

@article{G4DNA3,
	Author = {Bernal, M. A. and Bordage, M. C. and Brown, J. M. C. and Dav{\'\i}dkov{\'a}, M. and Delage, E. and El Bitar, Z. and Enger, S. A. and Francis, Z. and Guatelli, S. and Ivanchenko, V. N. and Karamitros, M. and Kyriakou, I. and Maigne, L. and Meylan, S. and Murakami, K. and Okada, S. and Payno, H. and Perrot, Y. and Petrovic, I. and Pham, Q. T. and Ristic-Fira, A. and Sasaki, T. and {\v S}t{\v e}p{\'a}n, V. and Tran, H. N. and Villagrasa, C. and Incerti, S.},
	Booktitle = {Physica Medica: European Journal of Medical Physics},
	Date = {2015/12/01},
	Doi = {10.1016/j.ejmp.2015.10.087},
	Isbn = {1120-1797},
	Journal = {Physica Medica: European Journal of Medical Physics},
	M3 = {doi: 10.1016/j.ejmp.2015.10.087},
	Month = {2024/10/11},
	Number = {8},
	Pages = {861--874},
	Publisher = {Elsevier},
	Title = {Track structure modeling in liquid water: A review of the Geant4-DNA very low energy extension of the Geant4 Monte Carlo simulation toolkit},
	Ty = {JOUR},
	Url = {https://doi.org/10.1016/j.ejmp.2015.10.087},
	Volume = {31},
	Year = {2015},
	Year1 = {2015},
	}

@article{G4DNA4,
author = {Incerti, S. and Ivanchenko, A. and Karamitros, M. and Mantero, A. and Moretto, P. and Tran, H. N. and Mascialino, B. and Champion, C. and Ivanchenko, V. N. and Bernal, M. A. and Francis, Z. and Villagrasa, C. and Baldacchino, G. and Guèye, P. and Capra, R. and Nieminen, P. and Zacharatou, C.},
title = {Comparison of GEANT4 very low energy cross section models with experimental data in water},
journal = {Medical Physics},
volume = {37},
number = {9},
pages = {4692-4708},
doi = {https://doi.org/10.1118/1.3476457},
url = {https://aapm.onlinelibrary.wiley.com/doi/abs/10.1118/1.3476457},
eprint = {https://aapm.onlinelibrary.wiley.com/doi/pdf/10.1118/1.3476457},
year = {2010}
}

@article{G4DNA5,
author = {Incerti, S. and Baldacchino, G. and Bernal, M. and Capra, R. and Champion, C. and Francis, Z. and Gu\`{e}ye, P. and Mantero, A. and Mascialino, B. and Moretto, P. and Nieminen, P. and Villagrasa, C. and Zacharatou, C.},
title = {The Geant4-DNA project},
journal = {International Journal of Modeling, Simulation, and Scientific Computing},
volume = {01},
number = {02},
pages = {157-178},
year = {2010},
doi = {10.1142/S1793962310000122},
URL = {https://doi.org/10.1142/S1793962310000122},
eprint = {https://doi.org/10.1142/S1793962310000122},
}

@article{kempf2025,
title = {Diffusion and mobility measurements for propane gas with a nanodosimetric detector},
journal = {Radiation Physics and Chemistry},
volume = {226},
pages = {112274},
year = {2025},
issn = {0969-806X},
doi = {https://doi.org/10.1016/j.radphyschem.2024.112274},
url = {https://www.sciencedirect.com/science/article/pii/S0969806X24007667},
author = {Irina Kempf and Uwe Schneider}
}

@Article{conte2023,
AUTHOR = {Conte, Valeria and Bianchi, Anna and Selva, Anna},
TITLE = {Track Structure of Light Ions: The Link to Radiobiology},
JOURNAL = {International Journal of Molecular Sciences},
VOLUME = {24},
YEAR = {2023},
NUMBER = {6},
ARTICLE-NUMBER = {5826},
URL = {https://www.mdpi.com/1422-0067/24/6/5826},
PubMedID = {36982899},
ISSN = {1422-0067},
DOI = {10.3390/ijms24065826}
}

@article{geant4_1,
title = {Recent developments in Geant4},
journal = {Nuclear Instruments and Methods in Physics Research Section A: Accelerators, Spectrometers, Detectors and Associated Equipment},
volume = {835},
pages = {186-225},
year = {2016},
issn = {0168-9002},
doi = {https://doi.org/10.1016/j.nima.2016.06.125},
url = {https://www.sciencedirect.com/science/article/pii/S0168900216306957},
author = {J. Allison and K. Amako and J. Apostolakis and P. Arce and M. Asai and T. Aso and E. Bagli and A. Bagulya and S. Banerjee and G. Barrand and B.R. Beck and A.G. Bogdanov and D. Brandt and J.M.C. Brown and H. Burkhardt and Ph. Canal and D. Cano-Ott and S. Chauvie and K. Cho and G.A.P. Cirrone and G. Cooperman and M.A. Cortés-Giraldo and G. Cosmo and G. Cuttone and G. Depaola and L. Desorgher and X. Dong and A. Dotti and V.D. Elvira and G. Folger and Z. Francis and A. Galoyan and L. Garnier and M. Gayer and K.L. Genser and V.M. Grichine and S. Guatelli and P. Guèye and P. Gumplinger and A.S. Howard and I. Hřivnáčová and S. Hwang and S. Incerti and A. Ivanchenko and V.N. Ivanchenko and F.W. Jones and S.Y. Jun and P. Kaitaniemi and N. Karakatsanis and M. Karamitros and M. Kelsey and A. Kimura and T. Koi and H. Kurashige and A. Lechner and S.B. Lee and F. Longo and M. Maire and D. Mancusi and A. Mantero and E. Mendoza and B. Morgan and K. Murakami and T. Nikitina and L. Pandola and P. Paprocki and J. Perl and I. Petrović and M.G. Pia and W. Pokorski and J.M. Quesada and M. Raine and M.A. Reis and A. Ribon and A. {Ristić Fira} and F. Romano and G. Russo and G. Santin and T. Sasaki and D. Sawkey and J.I. Shin and I.I. Strakovsky and A. Taborda and S. Tanaka and B. Tomé and T. Toshito and H.N. Tran and P.R. Truscott and L. Urban and V. Uzhinsky and J.M. Verbeke and M. Verderi and B.L. Wendt and H. Wenzel and D.H. Wright and D.M. Wright and T. Yamashita and J. Yarba and H. Yoshida}
}

@ARTICLE{geant4_2,
  author={Allison, J. and Amako, K. and Apostolakis, J. and Araujo, H. and Arce Dubois, P. and Asai, M. and Barrand, G. and Capra, R. and Chauvie, S. and Chytracek, R. and Cirrone, G.A.P. and Cooperman, G. and Cosmo, G. and Cuttone, G. and Daquino, G.G. and Donszelmann, M. and Dressel, M. and Folger, G. and Foppiano, F. and Generowicz, J. and Grichine, V. and Guatelli, S. and Gumplinger, P. and Heikkinen, A. and Hrivnacova, I. and Howard, A. and Incerti, S. and Ivanchenko, V. and Johnson, T. and Jones, F. and Koi, T. and Kokoulin, R. and Kossov, M. and Kurashige, H. and Lara, V. and Larsson, S. and Lei, F. and Link, O. and Longo, F. and Maire, M. and Mantero, A. and Mascialino, B. and McLaren, I. and Mendez Lorenzo, P. and Minamimoto, K. and Murakami, K. and Nieminen, P. and Pandola, L. and Parlati, S. and Peralta, L. and Perl, J. and Pfeiffer, A. and Pia, M.G. and Ribon, A. and Rodrigues, P. and Russo, G. and Sadilov, S. and Santin, G. and Sasaki, T. and Smith, D. and Starkov, N. and Tanaka, S. and Tcherniaev, E. and Tome, B. and Trindade, A. and Truscott, P. and Urban, L. and Verderi, M. and Walkden, A. and Wellisch, J.P. and Williams, D.C. and Wright, D. and Yoshida, H.},
  journal={IEEE Transactions on Nuclear Science}, 
  title={Geant4 developments and applications}, 
  year={2006},
  volume={53},
  number={1},
  pages={270-278},
  doi={10.1109/TNS.2006.869826}}

@article{geant4_3,
title = {Geant4—a simulation toolkit},
journal = {Nuclear Instruments and Methods in Physics Research Section A: Accelerators, Spectrometers, Detectors and Associated Equipment},
volume = {506},
number = {3},
pages = {250-303},
year = {2003},
issn = {0168-9002},
doi = {https://doi.org/10.1016/S0168-9002(03)01368-8},
url = {https://www.sciencedirect.com/science/article/pii/S0168900203013688},
author = {S. Agostinelli and J. Allison and K. Amako and J. Apostolakis and H. Araujo and P. Arce and M. Asai and D. Axen and S. Banerjee and G. Barrand and F. Behner and L. Bellagamba and J. Boudreau and L. Broglia and A. Brunengo and H. Burkhardt and S. Chauvie and J. Chuma and R. Chytracek and G. Cooperman and G. Cosmo and P. Degtyarenko and A. Dell'Acqua and G. Depaola and D. Dietrich and R. Enami and A. Feliciello and C. Ferguson and H. Fesefeldt and G. Folger and F. Foppiano and A. Forti and S. Garelli and S. Giani and R. Giannitrapani and D. Gibin and J.J. {Gómez Cadenas} and I. González and G. {Gracia Abril} and G. Greeniaus and W. Greiner and V. Grichine and A. Grossheim and S. Guatelli and P. Gumplinger and R. Hamatsu and K. Hashimoto and H. Hasui and A. Heikkinen and A. Howard and V. Ivanchenko and A. Johnson and F.W. Jones and J. Kallenbach and N. Kanaya and M. Kawabata and Y. Kawabata and M. Kawaguti and S. Kelner and P. Kent and A. Kimura and T. Kodama and R. Kokoulin and M. Kossov and H. Kurashige and E. Lamanna and T. Lampén and V. Lara and V. Lefebure and F. Lei and M. Liendl and W. Lockman and F. Longo and S. Magni and M. Maire and E. Medernach and K. Minamimoto and P. {Mora de Freitas} and Y. Morita and K. Murakami and M. Nagamatu and R. Nartallo and P. Nieminen and T. Nishimura and K. Ohtsubo and M. Okamura and S. O'Neale and Y. Oohata and K. Paech and J. Perl and A. Pfeiffer and M.G. Pia and F. Ranjard and A. Rybin and S. Sadilov and E. {Di Salvo} and G. Santin and T. Sasaki and N. Savvas and Y. Sawada and S. Scherer and S. Sei and V. Sirotenko and D. Smith and N. Starkov and H. Stoecker and J. Sulkimo and M. Takahata and S. Tanaka and E. Tcherniaev and E. {Safai Tehrani} and M. Tropeano and P. Truscott and H. Uno and L. Urban and P. Urban and M. Verderi and A. Walkden and W. Wander and H. Weber and J.P. Wellisch and T. Wenaus and D.C. Williams and D. Wright and T. Yamada and H. Yoshida and D. Zschiesche}
}

@article{pietrzak2018,
  author    = {Pietrzak, M. and Pszona, S. and Bantsar, A.},
  title     = {Measurements of Spatial Correlations of Ionisation Clusters in the Track of Carbon Ions-First Results},
  journal   = {Radiation Protection Dosimetry},
  volume    = {180},
  number    = {1-4},
  pages     = {162--167},
  year      = {2018},
  doi       = {10.1093/rpd/ncy079},
  url       = {https://doi.org/10.1093/rpd/ncy079}
}

@article{bancer2020,
title = {Particle track structure measurements from 0.5 to 18 nm in nitrogen using the Jet Counter nanodosemeter},
journal = {Radiation Physics and Chemistry},
volume = {172},
pages = {108805},
year = {2020},
issn = {0969-806X},
doi = {https://doi.org/10.1016/j.radphyschem.2020.108805},
url = {https://www.sciencedirect.com/science/article/pii/S0969806X19308126},
author = {Aleksandr Bancer and Marcin Pietrzak and Monika Mietelska}
}

@article{merza2025,
title = {Garfield++ and Geant4-DNA simulation of a compact THGEM-based nanodosimeter},
journal = {Nuclear Instruments and Methods in Physics Research Section A: Accelerators, Spectrometers, Detectors and Associated Equipment},
volume = {1080},
pages = {170729},
year = {2025},
issn = {0168-9002},
doi = {https://doi.org/10.1016/j.nima.2025.170729},
url = {https://www.sciencedirect.com/science/article/pii/S0168900225005303},
author = {Victor Merza and Aleksandr Bancer and Ana Belchior and Beata Brzozowska and João F. Canhoto and Khaled Katmeh and Marcin Pietrzak and Antoni Ruciński and Reinhard Schulte}
}

@article{kempf20252,
title = {Development and characterization of a compact nanodosimetric detector},
journal = {Nuclear Instruments and Methods in Physics Research Section A: Accelerators, Spectrometers, Detectors and Associated Equipment},
volume = {1075},
pages = {170337},
year = {2025},
issn = {0168-9002},
doi = {https://doi.org/10.1016/j.nima.2025.170337},
url = {https://www.sciencedirect.com/science/article/pii/S016890022500138X},
author = {Irina Kempf and Tamara Melina Hoffmann and Jürgen Besserer and Uwe Schneider}
}

@article{wang2010,
title = {Development of multi-gap resistive plate chambers with low-resistive silicate glass electrodes for operation at high particle fluxes and large transported charges},
journal = {Nuclear Instruments and Methods in Physics Research Section A: Accelerators, Spectrometers, Detectors and Associated Equipment},
volume = {621},
number = {1},
pages = {151-156},
year = {2010},
issn = {0168-9002},
doi = {https://doi.org/10.1016/j.nima.2010.04.056},
url = {https://www.sciencedirect.com/science/article/pii/S0168900210009058},
author = {Jingbo Wang and Yi Wang and Xianglei Zhu and Weicheng Ding and Yuanjing Li and Jianping Cheng and Nobert Herrmann and Ingo Deppner and Yapeng Zhang and P. Loizeau and P. Senger and D. Gonzalez-Diaz}
}

@article{wang2019,
doi = {10.1088/1748-0221/14/06/C06015},
url = {https://doi.org/10.1088/1748-0221/14/06/C06015},
year = {2019},
month = {jun},
publisher = {},
volume = {14},
number = {06},
pages = {C06015},
author = {Wang, Y. and Zhang, Q. and Han, D. and Wang, F. and Yu, Y. and Lyu, P. and Li, Y.},
title = {Status of technology of MRPC time of flight system},
journal = {Journal of Instrumentation}
}

@article{Arazi2018,
doi = {10.1088/1742-6596/1029/1/012004},
url = {https://doi.org/10.1088/1742-6596/1029/1/012004},
year = {2018},
month = {may},
publisher = {IOP Publishing},
volume = {1029},
number = {1},
pages = {012004},
author = {Arazi, Lior},
title = {On the possibility of positive-ion detection in gaseous TPCs and its potential use for neutrinoless double beta decay searches in 136Xe},
journal = {Journal of Physics: Conference Series}
}

@article{Vance1968,
  title = {Auger Electron Emission from Clean Mo Bombarded by Positive Ions. II. Effect of Angle of Incidence},
  author = {Vance, Dennis W.},
  journal = {Phys. Rev.},
  volume = {169},
  issue = {2},
  pages = {252--262},
  numpages = {0},
  year = {1968},
  month = {May},
  publisher = {American Physical Society},
  doi = {10.1103/PhysRev.169.252},
  url = {https://link.aps.org/doi/10.1103/PhysRev.169.252}
}

@article{Vance1968_2,
  title = {Auger Electron Emission from Clean Mo Bombarded by Positive Ions. III. Effect of Electronically Excited Ions},
  author = {Vance, Dennis W.},
  journal = {Phys. Rev.},
  volume = {169},
  issue = {2},
  pages = {263--272},
  numpages = {0},
  year = {1968},
  month = {May},
  publisher = {American Physical Society},
  doi = {10.1103/PhysRev.169.263},
  url = {https://link.aps.org/doi/10.1103/PhysRev.169.263}
}

@article{Elsbergen2000,
author = {van Elsbergen, Volker and Bachmann, Peter K. and Juestel, Thomas},
title = {16.3: Ion-Induced Secondary Electron Emission: A Comparative Study},
journal = {SID Symposium Digest of Technical Papers},
volume = {31},
number = {1},
pages = {220-223},
doi = {https://doi.org/10.1889/1.1832922},
url = {https://sid.onlinelibrary.wiley.com/doi/abs/10.1889/1.1832922},
eprint = {https://sid.onlinelibrary.wiley.com/doi/pdf/10.1889/1.1832922},
year = {2000}
}

@article{Kim2000,
author = {Kim, Dae-Il and Lim, Young-Guon and Kim, Young-Guon and Ko, Jae-Jun and Lee, Chon Wo and Cho, Guang-Sup and Choi, Eun-Ha},
title = {Ion-Induced Secondary Electron Emission Coefficient ($\gamma$) of Bulk MgO-Single Crystals},
journal = {Jpn. J. Appl. Phys.},
volume = {39},
number = {4A},
pages = {1890-1891},
url = {https://iopscience.iop.org/article/10.1143/JJAP.39.1890/pdf},
year = {2000}
}

@techreport{CERN_Tech_Note,
    author = {Barnard, J and Bojko, I and Hilleret, N},
    title = {Measurements of the Secondary Electron Emission of Some Insulators},
    institution = {European Laboratory for Particle Physics CERN-LHC Division},
    year = {1997},
    url = {https://cds.cern.ch/record/1514931/files/arxiv:1302.2333.pdf},
}

@manual{GUM,
    title = {Guide to the expression of uncertainty in measurement},
    url = {https://www.bipm.org/documents/20126/194484570/JCGM_GUM-1/74e7aa56-2403-7037-f975-cd6b555b80e6},
    author = {JCGM},
    organization = {Bureau International des Poids et Mesures (BIPM)},
    year = {2023}
}

@article{Bug2013,
  title = {Ionization cross section data of nitrogen, methane, and propane for light ions and electrons and their suitability for use in track structure simulations},
  author = {Bug, Marion U. and Gargioni, Elisabetta and Nettelbeck, Heidi and Baek, Woon Yong and Hilgers, Gerhard and Rosenfeld, Anatoly B. and Rabus, Hans},
  journal = {Phys. Rev. E},
  volume = {88},
  issue = {4},
  pages = {043308},
  numpages = {21},
  year = {2013},
  month = {Oct},
  publisher = {American Physical Society},
  doi = {10.1103/PhysRevE.88.043308},
  url = {https://link.aps.org/doi/10.1103/PhysRevE.88.043308}
}

@article{heylen1975,
author = {A. E. D. Heylen},
title = {Ionization coefficients and sparking voltages from methane to butane},
journal = {International Journal of Electronics},
volume = {39},
number = {6},
pages = {653--660},
year = {1975},
publisher = {Taylor \& Francis},
doi = {10.1080/00207217508920532},
URL = {https://doi.org/10.1080/00207217508920532},
eprint = {https://doi.org/10.1080/00207217508920532}
}

@Article{mietelska2024,
AUTHOR = {Mietelska, Monika and Pietrzak, Marcin and Bancer, Aleksandr and Ruciński, Antoni and Szefliński, Zygmunt and Brzozowska, Beata},
TITLE = {Ionization Detail Parameters for DNA Damage Evaluation in Charged Particle Radiotherapy: Simulation Study Based on Cell Survival Database},
JOURNAL = {International Journal of Molecular Sciences},
VOLUME = {25},
YEAR = {2024},
NUMBER = {10},
ARTICLE-NUMBER = {5094},
URL = {https://www.mdpi.com/1422-0067/25/10/5094},
PubMedID = {38791135},
ISSN = {1422-0067},
DOI = {10.3390/ijms25105094}
}

@article{pak1992,
    author = {Pak, Hoyoung and Kushner, Mark J.},
    title = {Breakdown characteristics in nonplanar geometries and hollow cathode pseudospark switches},
    journal = {Journal of Applied Physics},
    volume = {71},
    number = {1},
    pages = {94-100},
    year = {1992},
    month = {01},
    issn = {0021-8979},
    doi = {10.1063/1.350653},
    url = {https://doi.org/10.1063/1.350653}
}

@incollection{Varga1992,
  author    = {Varga, P. and Winter, H.},
  title     = {Slow particle-induced electron emission from solid surfaces},
  booktitle = {Particle Induced Electron Emission II},
  publisher = {Springer Berlin Heidelberg},
  address   = {Berlin, Heidelberg},
  year      = {1992},
  pages     = {149--214},
  isbn      = {978-3-540-47447-0},
  doi       = {10.1007/BFb0038300},
  url       = {https://doi.org/10.1007/BFb0038300}
}

@misc{merza2026,
      title={Toward Ionization Cluster Size Measurements with a Compact Nanodosimeter}, 
      author={Victor Merza and Aleksandr Bancer and Vladimir Bashkirov and Ana Belchior and Beata Brzozowska and Piotr Gasik and Jaroslaw Grzyb and Khaled Katmeh and Marcin Pietrzak and Antoni Ruciński and Reinhard Schulte},
      year={2026},
      eprint={2608.29648},
      archivePrefix={arXiv},
      primaryClass={physics.ins-det},
      url={https://arxiv.org/abs/2608.29648}, 
}

@article{conte2010,
title = {First track-structure measurements of 20MeV protons with the STARTRACK apparatus},
journal = {Radiation Measurements},
volume = {45},
number = {10},
pages = {1213-1216},
year = {2010},
note = {PROCEEDINGS OF THE 11TH SYMPOSIUM ON NEUTRON AND ION DOSIMETRY},
issn = {1350-4487},
doi = {https://doi.org/10.1016/j.radmeas.2010.04.004},
url = {https://www.sciencedirect.com/science/article/pii/S1350448710001034},
author = {V. Conte and L. {De Nardo} and P. Colautti and A. Ferretti and B. Grosswendt and M. Lombardi and M. Poggi and S. Canella and D. Moro and G. Tornielli}
}

@article{conte2012,
doi = {10.1088/1367-2630/14/9/093010},
url = {https://doi.org/10.1088/1367-2630/14/9/093010},
year = {2012},
month = {sep},
publisher = {IOP Publishing},
volume = {14},
number = {9},
pages = {093010},
author = {Conte, V and Colautti, P and Grosswendt, B and Moro, D and De Nardo, L},
title = {Track structure of light ions: experiments and simulations},
journal = {New Journal of Physics}
}

@article{denardo2007,
    author = {DeNardo, L. and Conte, V. and Poggi, M. and Canella, S. and Colautti, P. and Moro, D. and Tornielli, G.},
    title = {The Startrack experiment},
    journal = {Radiation Protection Dosimetry},
    volume = {126},
    number = {1-4},
    pages = {453-456},
    year = {2007},
    month = {08},
    issn = {0144-8420},
    doi = {10.1093/rpd/ncm091},
    url = {https://doi.org/10.1093/rpd/ncm091},
    eprint = {https://academic.oup.com/rpd/article-pdf/126/1-4/453/4565515/ncm091.pdf},
}

@article{conte2011,
    author = {Conte, V. and Colautti, P. and De Nardo, L. and Ferretti, A. and Poggi, M. and Moro, D. and Lombardi, M. and Tornielli, G. and Grosswendt, B.},
    title = {Track nanodosimetry of 20-MeV protons at 20 nm},
    journal = {Radiation Protection Dosimetry},
    volume = {143},
    number = {2-4},
    pages = {455-458},
    year = {2011},
    month = {02},
    issn = {0144-8420},
    doi = {10.1093/rpd/ncq402},
    url = {https://doi.org/10.1093/rpd/ncq402},
    eprint = {https://academic.oup.com/rpd/article-pdf/143/2-4/455/4555285/ncq402.pdf},
}

@article{Grosswendt2014,
    author = {Grosswendt, B. and Conte, V. and Colautti, P.},
    title = {An upgraded track structure model: experimental validation},
    journal = {Radiation Protection Dosimetry},
    volume = {161},
    number = {1-4},
    pages = {464-468},
    year = {2014},
    month = {10},
    issn = {0144-8420},
    doi = {10.1093/rpd/nct322},
    url = {https://doi.org/10.1093/rpd/nct322},
    eprint = {https://academic.oup.com/rpd/article-pdf/161/1-4/464/9583465/nct322.pdf},
}

@article{selva2020,
title = {Nanodosimetry of light ions in targets of different size},
journal = {Radiation Physics and Chemistry},
volume = {172},
pages = {108806},
year = {2020},
issn = {0969-806X},
doi = {https://doi.org/10.1016/j.radphyschem.2020.108806},
url = {https://www.sciencedirect.com/science/article/pii/S0969806X19309740},
author = {A. Selva and P. Colautti and V. Conte}
}

@article{conte2014,
    author = {Conte, V. and Colautti, P. and Moro, D. and Grosswendt, B.},
    title = {Track structure of carbon ions: measurements and simulations},
    journal = {Radiation Protection Dosimetry},
    volume = {161},
    number = {1-4},
    pages = {445-448},
    year = {2014},
    month = {10},
    issn = {0144-8420},
    doi = {10.1093/rpd/nct266},
    url = {https://doi.org/10.1093/rpd/nct266},
    eprint = {https://academic.oup.com/rpd/article-pdf/161/1-4/445/9582612/nct266.pdf},
}

@article{denardo2002b,
    author = {De Nardo, L. and Colautti, P. and Y. Baek, W. and Grosswendt, B. and Alkaa, A. and Ségur, P. and Tornielli, G.},
    title = {Track Nanodosimetry of an Alpha Particle},
    journal = {Radiation Protection Dosimetry},
    volume = {99},
    number = {1-4},
    pages = {355-358},
    year = {2002},
    month = {06},
    issn = {0144-8420},
    doi = {10.1093/oxfordjournals.rpd.a006801},
    url = {https://doi.org/10.1093/oxfordjournals.rpd.a006801},
    eprint = {https://academic.oup.com/rpd/article-pdf/99/1-4/355/9927772/355.pdf},
}

@article{garty2002b,
    author = {Garty, G. and Shchemelinin, S. and Breskin, A. and Chechik, R. and Orion, I. and P. Guedes, G. and Schulte, R. and Bashkirov, V. and Grosswendt, B.},
    title = {Wall-less Ion-counting Nanodosimetry Applied to Protons},
    journal = {Radiation Protection Dosimetry},
    volume = {99},
    number = {1-4},
    pages = {325-330},
    year = {2002},
    month = {06},
    issn = {0144-8420},
    doi = {10.1093/oxfordjournals.rpd.a006794},
    url = {https://doi.org/10.1093/oxfordjournals.rpd.a006794},
    eprint = {https://academic.oup.com/rpd/article-pdf/99/1-4/325/9927765/325.pdf},
}

@article{bashkirov2006,
    author = {Bashkirov, V. and Schulte, R. and Breskin, A. and Chechik, R. and Schemelinin, S. and Garty, G. and Wroe, A. and Sadrozinski, H. and Grosswendt, B.},
    title = {Ion-counting nanodosemeter with particle tracking capabilities},
    journal = {Radiation Protection Dosimetry},
    volume = {122},
    number = {1-4},
    pages = {415-419},
    year = {2006},
    month = {12},
    issn = {0144-8420},
    doi = {10.1093/rpd/ncl470},
    url = {https://doi.org/10.1093/rpd/ncl470},
    eprint = {https://academic.oup.com/rpd/article-pdf/122/1-4/415/9582179/ncl470.pdf},
}

@article{bronic_1999,
doi = {10.1088/0022-3727/32/24/314},
url = {https://doi.org/10.1088/0022-3727/32/24/314},
year = {1999},
month = {dec},
publisher = {},
volume = {32},
number = {24},
pages = {3179},
author = {Ines Krajcar Bronic and Bernd Grosswendt},
title = {Ionization 
coefficient in propane, propane-based tissue equivalent and 
dimethyl-ether in strong non-uniform electric fields},
journal = {Journal of Physics D: Applied Physics}
}

@Inbook{Hilke2020,
author="Hilke, H. J.
and Riegler, W.",
title="Gaseous Detectors",
bookTitle="Particle Physics Reference Library: Volume 2: Detectors for Particles and Radiation",
year="2020",
publisher="Springer International Publishing",
address="Cham",
chapter="4",
pages="91--136",
isbn="978-3-030-35318-6",
doi="10.1007/978-3-030-35318-6_4",
url="https://doi.org/10.1007/978-3-030-35318-6_4"
}

\end{document}